\documentclass[twocolumn,twocolappendix]{aastex63}
\usepackage{hyperref}
\usepackage[flushleft]{threeparttable}
\usepackage{placeins}
\usepackage{ulem}
\usepackage{natbib}
\shortauthors{Tran et al.}
\graphicspath{{./}{figures/}}

\setcitestyle{citesep={;}}
\setcitestyle{aysep={}}

\begin{document}

\title{Distances to Local Group Galaxies via Population II, Stellar Distance Indicators\\ I: The Sculptor Dwarf Spheroidal
\footnote{Based in part on observations made with the NASA/ESA \textit{Hubble Space Telescope}, obtained at the Space Telescope Science Institute, which is operated by the Association of Universities for Research in Astronomy, Inc. under NASA contract NAS 5-26555. These observations are associated with program No. 13691. Additional observations are credited to the Observatories of the Carnegie Institution of Science for the use of Magellan-Baade IMACS.}}

\correspondingauthor{Quang H. Tran}

\author[0000-0001-6532-6755]{Quang H. Tran}
\email{quangtran@utexas.edu}
\affiliation{Department of Astronomy, The University of Texas at Austin, 2515 Speedway, Stop C1400, Austin, TX 78712, USA}

\author[0000-0001-9664-0560]{Taylor J. Hoyt}
\affiliation{Department of Astronomy \& Astrophysics, University of Chicago, 5640 South Ellis Avenue, Chicago, IL 60637, USA}

\author[0000-0003-3431-9135]{Wendy L. Freedman}
\affiliation{Department of Astronomy \& Astrophysics, University of Chicago, 5640 South Ellis Avenue, Chicago, IL 60637, USA}

\author[0000-0002-1576-1676]{Barry F. Madore}
\affiliation{Department of Astronomy \& Astrophysics, University of Chicago, 5640 South Ellis Avenue, Chicago, IL 60637, USA}
\affiliation{The Observatories of the Carnegie Institution for Science, 813 Santa Barbara Street, Pasadena, CA 91101, USA}

\author[0000-0002-0119-1115]{Elias K. Oakes}
\affiliation{Department of Astronomy \& Astrophysics, University of Chicago, 5640 South Ellis Avenue, Chicago, IL 60637, USA}
\affiliation{Department of Physics, University of Connecticut, 196A Auditorium Road, Storrs, CT 06269, USA}

\author[0000-0003-1697-7062]{William Cerny}
\affiliation{Department of Astronomy \& Astrophysics, University of Chicago, 5640 South Ellis Avenue, Chicago, IL 60637, USA}

\author[0000-0003-2767-2379]{Dylan Hatt}
\affiliation{The Observatories of the Carnegie Institution for Science, 813 Santa Barbara Street, Pasadena, CA 91101, USA}

\author[0000-0002-1691-8217]{Rachael L. Beaton}
\affiliation{Department of Astrophysical Sciences, Princeton University, 4 Ivy Lane, Princeton, NJ 08544, USA}
\affiliation{The Observatories of the Carnegie Institution for Science, 813 Santa Barbara Street, Pasadena, CA 91101, USA}

\accepted{2022 May 13}
\submitjournal{ApJ}

\begin{abstract}
We determine the distance to the Sculptor Dwarf Spheroidal via three Population II stellar distance indicators: (a) the Tip of the Red Giant Branch (TRGB), (b) RR Lyrae variables (RRLs), and (c) the ridgeline of the blue horizontal branch (HB). High signal-to-noise, wide-field $VI$ imaging that covers an area $48' \times 48'$ and reaches a photometric depth approximately 2~mag fainter than the HB was acquired with the Magellan-Baade 6.5m telescope. The true modulus derived from Sculptor's TRGB is found to be $\mu^\mathrm{TRGB}_o = 19.59 \pm 0.07_\mathrm{stat} \pm 0.05_\mathrm{sys}$ mag. Along with periods adopted from the literature, newly acquired RRL phase points are fit with template light curves to determine $\mu_{W_{I,V-I}}^\mathrm{RRL} = 19.60 \pm 0.01_\mathrm{stat} \pm 0.05_\mathrm{sys}$ mag. Finally, the HB distance is found to be $\mu^\mathrm{HB}_o = 19.54 \pm 0.03_\mathrm{stat} \pm 0.09_\mathrm{sys}$ mag. Absolute calibrations of each method are anchored by independent geometric zero-points, utilizes a different class of stars, and are determined from the same photometric calibration.
\end{abstract}

\keywords{distance scale --- galaxies: individual (Sculptor) --- stars: Population II --- stars: variables: RR Lyrae}

\section{\textbf{Introduction} \label{sec:intro}}

Dwarf spheroidal galaxies (dSph) in the Local Group are excellent laboratories within which Population II\footnote{Population II here refers to old-age, metal-poor stars.} (Population II) methods of distance measurement can be tested. Due to their proximity to the Milky Way, many of them are excellent targets for both ground- and space-based distance determinations. Furthermore, dwarf galaxies in the Local Group are well populated by older stellar components \citep{Tolstoy2009}. We leverage these advantages to derive precise distances to two nearby Local Group dwarf galaxies, the Sculptor dSph in this paper and the Fornax dSph in a companion paper \citep{Oakes2022}.

As the first Milky Way dwarf spheroidal satellite ever discovered \citep{Shapley1938}, the Sculptor Dwarf spheroidal is a well-characterized galaxy in the Local Group. Sculptor has been observed to have multiple stellar components and a spread in metallicity \citep{deBoer2011}. \citet{Tolstoy2004} were able to identify two stellar components of stars, a ``metal-poor'' component ([Fe/H] $<$ -1.7~dex) and a ``metal-rich'' component ([Fe/H] $>$ -1.7~dex). Similarly, \citet{Clementini2005}, using low resolution spectra of 107 RR Lyrae (RRLs), found a mean metallicity of $-1.83 \pm 0.03$ dex for Sculptor, with a full range of $-2.4 <$ [Fe/H] $< 0.8$ dex reflecting this split in populations.

Most prior distances calculated for Sculptor have used RRLs \citep[e.g.,][]{Pietrzynski2008, Martinez-Vazquez2015, Garofalo2018}. In this work, we present new determinations of the distance to the Sculptor dSph using the TRGB, $VI$ period-luminosity relations of RRLs, and the horizontal branch (HB), allowing us to assess and confirm consistency amongst the three methods of Population II distance measurement.

The structure of this paper is as follows. \autoref{sec:Data} describes the observations and data used in this study. We present our distance measurement from the TRGB in \autoref{sec:trgb}, from RRLs in \autoref{sec:calibrate} and from the HB in \autoref{sec:zahb}. Comparisons with previous distance measurements are made in \autoref{sec:compare}. Finally, \autoref{sec:Conclusion} summarizes our results.

\section{\label{sec:Data} \textbf{Data}}
Observations, image processing, and photometry procedures follow the methods detailed in \citet{Hatt2017}. The data analyzed in this work come from both space- and ground-based imaging, as specified in \autoref{HST_IMACS}. Each image was reduced, with sources identified,  photometered, and calibrated as explained in \autoref{sec:Photo}.

\subsection{\label{HST_IMACS} IMACS and HST Data}
\begin{figure*}[ht!]
	\centering
	\includegraphics[width=0.9\textwidth]{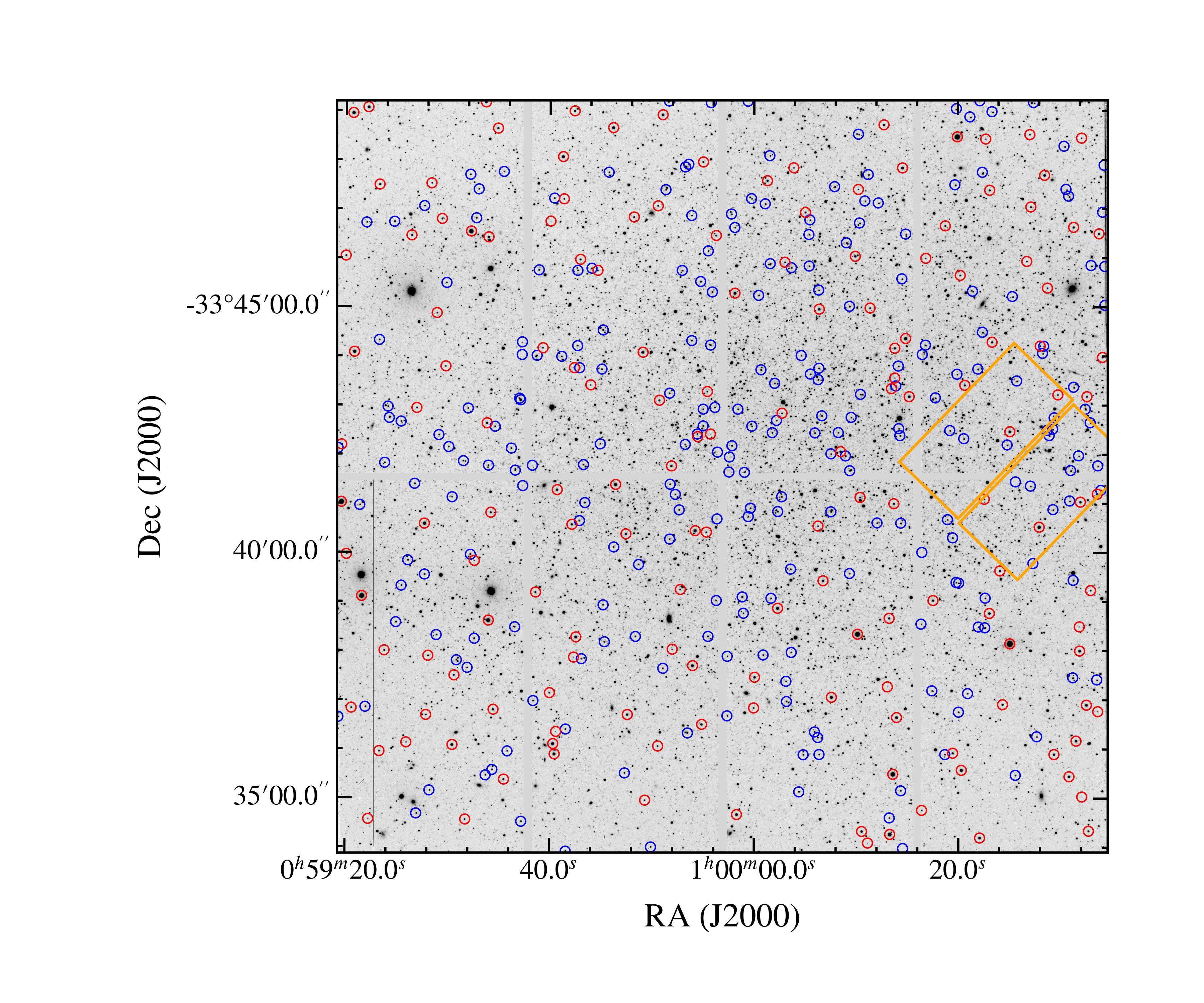}
	\caption{Representative mosaic (comprised of eight chips) image of the Sculptor Dwarf Spheroidal galaxy taken with the Magellan IMACS instrument in the \textit{V}-band. Nine fields of this size ($15^\prime.46 \times 15^\prime.46$) cover the inner $45'$ of Sculptor and, with at least two visits to each field, make up the imaging used in this study. \textit{HST} ACS/WFC F814W footprint is outlined in orange. Stetson standard stars are circled in red (see \autoref{sec:Photo}); RRLs are circled in blue (see \autoref{sec:RRL_Results}).}
	\label{fig:image}
\end{figure*}

As stars in the final stages of the Red Giant Branch (RGB) are short-lived, a precise measurement of the TRGB requires sampling a region of a galaxy large enough to contain a sufficient number of these particular stars, which will be adequate to fully populate the luminosity function (LF) at the discontinuity. Sculptor is sufficiently low-mass and extended on the sky ($\sim$1~deg$^2$) that a wide-field survey instrument is a requirement to accumulate a large enough sample of RGB stars. Over the course of two nights, 24 July, 2014 and 19 September, 2014, 17 overlapping pointings of Sculptor were taken with the Inamori-Magellan Areal Camera \& Spectrograph (IMACS) on the Magellan-Baade 6.5-meter telescope at Las Campanas Observatory (LCO) \citep{Dressler2011}. The IMACS instrument was used at f/4, providing a $15.46^\prime \times 15.46^\prime$ field of view with a $0.2^{\prime\prime}$ pixel$^{-1}$ plate scale. At each pointing, a series of short (10--30s) and long (300s) exposures were taken to ensure a dynamic range that captured both RR Lyrae stars and the TRGB, in both of the Johnson-Cousins $V$ and $I$ filters. The IMACS camera has 8 different chips for each exposure; they were individually debiased and flat-field-corrected. The calibrated individual chip exposures were then stitched together into a single eight-chip mosaic (See \autoref{fig:image}). WCS coordinates were added to each mosaic image using the \texttt{Astrometry.net}\footnote{\href{http://nova.astrometry.net/}{http://nova.astrometry.net/}} open source code \citep{Lang2010}.

In the \autoref{sec:appen_c}, we establish a direct tie-in between ground- and space-based photometric systems, using two 30s HST epochs with the ACS/WFC instrument in the F814W filter that were taken on 21 June, 2015 (PID: GO-13691, PI: Freedman) \citep{Freedman2014}. The ACS/WFC instrument has a $202^{\prime\prime} \times 202^{\prime\prime}$ field of view with a scale $0.05^{\prime\prime}$ pixel$^{-1}$. The \textit{flc} images used were flat-field and CTE corrected by the automated STScI processing pipeline. \autoref{fig:image} is an example of the IMACS coverage and the overlapping \textit{HST} ACS/WFC footprint.

\subsection{\label{sec:Photo} Photometry}
All photometry performed on the IMACS images used the DAOPHOT and ALLFRAME photometry software package \citep{Stetson1987, Stetson1994} following the methodology outlined in the DAOPHOT-II User Manual \citep{Stetson2000}. For every epoch, we obtained both short and long exposures. The longer exposures are deeper to maximize completeness at and below the HB. Short exposures were taken to guard against saturation effects at the TRGB. The final instrumental source catalog was created by replacing photometric measurements flagged as saturated in the long exposures with short exposure photometry. Saturation limits were identified in the distributions of \textit{Chi} (PSF goodness-of-fit) vs. magnitude, then confirmed visually in the images. A data quality cut was applied to each image based on the prescription from \citet{Beaton2019}, which elimated sources based on the DAOPHOT photometric error, \textit{Chi}, and sharpness parameters.

\textit{HST} images were reduced using TinyTim PSF models \citep{Krist2011} with an automated reduction pipeline developed by the Chicago-Carnegie Hubble Program \citep[CCHP;][]{Beaton2019}.

We next calibrated the IMACS instrumental magnitudes to the Stetson-Landolt standards\footnote{\href{http://www.cadc-ccda.hia-iha.nrc-cnrc.gc.ca/en/community/STETSON/standards/}{http://www.cadc-ccda.hia-iha.nrc-cnrc.gc.ca/en/community/STETSON/standards/}}. Overlapping sources between the IMACS photometry and Stetson standard stars were matched using their WCS coordinates, with a $0.015^\prime$ radius threshold. These overlapping sources were visually inspected to remove stars hit by cosmic rays, affected by saturation, near the edge of the CCD, or likely false matches. Matched sources in each frame were combined into a single catalog of calibration stars for the $V$- and $I$--bands.

\begin{figure*}[ht!]
    \centering
	\includegraphics[width=1.0\textwidth]{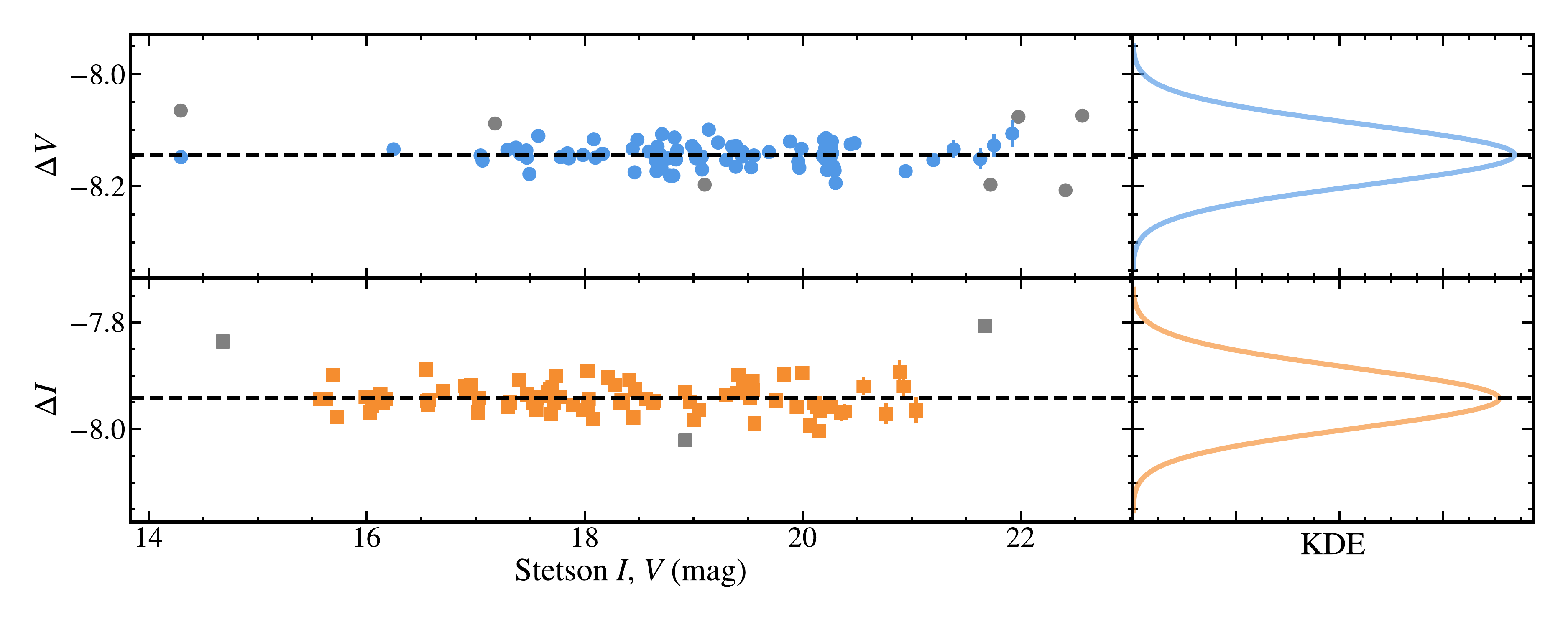}
	\includegraphics[width=0.98\textwidth]{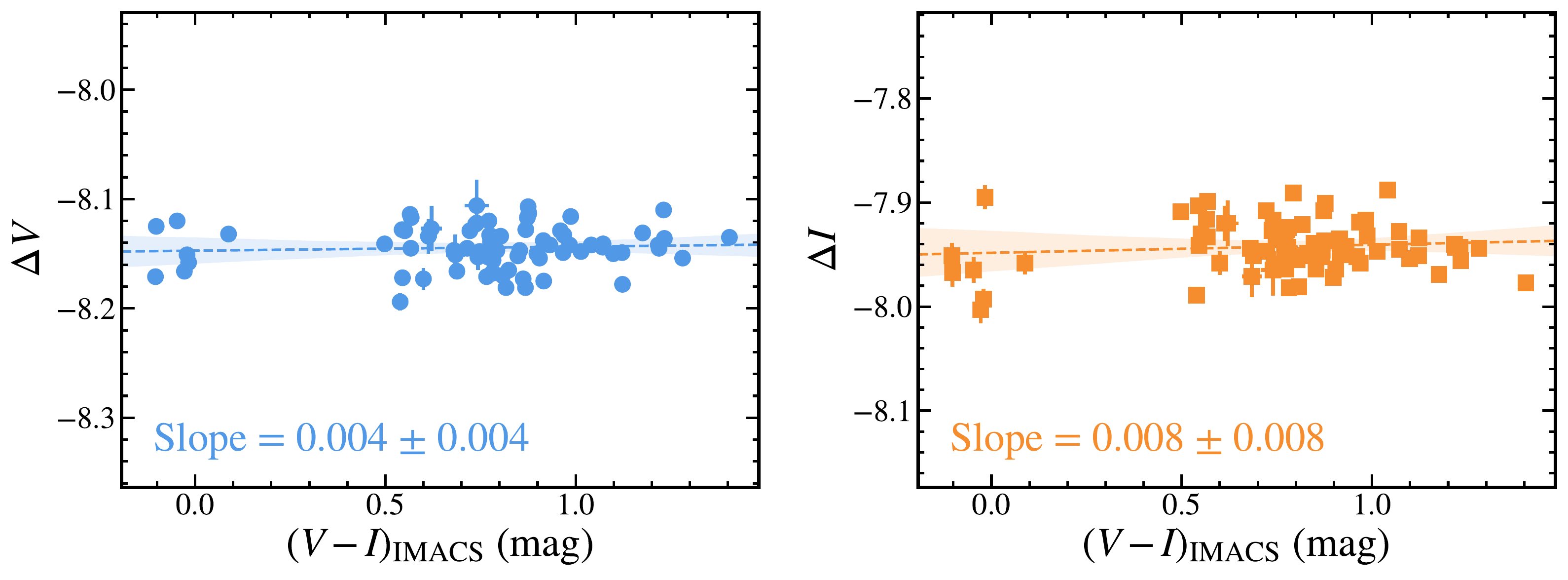}
	\caption{A calibration sample for IMACS $V$- (top, blue circles) and $I$--band (bottom, orange squares) to Stetson standards for a single, representative field. The final average magnitude offsets adopted are detailed in \autoref{tab:cali}. Magnitude offsets are the instrumental magnitudes minus the corresponding Stetson photometry. The gray points are sources 2$\sigma$ from the median and are excluded from the final offset mean value. A Gaussian kernel density estimate (KDE) of the sample is shown on the right. The photometric difference is plotted against IMACS $(V-I)$ color on the bottom. A linear fit is applied to the data and the 68\% confidence intervals are plotted. There is no evidence for a color term as the slopes of the fits are consistent with zero.}
	\label{fig:imacs_comp}%
\end{figure*}

The top of \autoref{fig:imacs_comp} displays the calibration from IMACS \textit{VI} to Stetson-Landolt \textit{VI} magnitudes for a representative image. Each composite mosaic offset was calculated from the mean of that frame's calibration and all stars that were 2$\sigma$ from the median (dotted line) have been excluded (shown as gray points). The average photometric zero-point calibrations for all filters and exposures are further detailed in \autoref{tab:cali}. We also investigate if there is a zero-point dependence on $(V-I)$ color and confirm that there is no presence of a color term in any of the images. The bottom of \autoref{fig:imacs_comp} shows the $VI$--band calibrations as a function of IMACS $(V-I)$ color for a representative image.

\begin{table}[t!]
    \caption{IMACS to Stetson-Landolt Offsets}
    \centering
    \label{tab:cali}
    \begin{tabular}{c c c c c}
    \toprule
    \small
    Filter & Exp. (s) & $\mathrm{IMACS}-\mathrm{Stetson}$ (mag) & $N_\mathrm{star}$ & $N_\mathrm{frame}$ \\
    \hline
    $V$ & 300 & $-8.16 \pm 0.05$ & 131 & 17 \\
    $I$ & 300 & $-7.98 \pm 0.05$ & 119 & 17 \\
    $V$ & 60 & $-6.44 \pm 0.02$ & 132 & 3\\
    $I$ & 60 & $-6.27 \pm 0.03$ & 136 & 3\\
    $V$ & 30 & $-5.64 \pm 0.04$ & 111 & 10 \\
    $I$ & 30 & $-5.45 \pm 0.03$ & 92 & 11 \\
    $V$ & 10 & $-4.530 \pm 0.003$ & 122 & 3 \\
    $I$ & 10 & $-4.315 \pm 0.003$ & 106 & 2 \\
    \hline
    \end{tabular}
\end{table}

It has historically been assumed that internal reddening is negligible for Sculptor \citep[e.g.,][]{Pietrzynski2008, Martinez-Vazquez2015, Garofalo2018}, and that has been shown to be an accurate assumption \citep{Putman_2021}. We deredden the photometric catalogs with a foreground reddening $E(B - V) = 0.016$ mag \citep{Schlegel1998, Schlafly2011}, or $A_I = 0.027$ mag and $A_V = 0.050$ mag assuming a \citet{Fitzpatrick1999} reddening law and $R_V = 3.1$.

\begin{figure*}[ht!]%
\onecolumngrid
	\centering
	\includegraphics[width=0.8\textwidth]{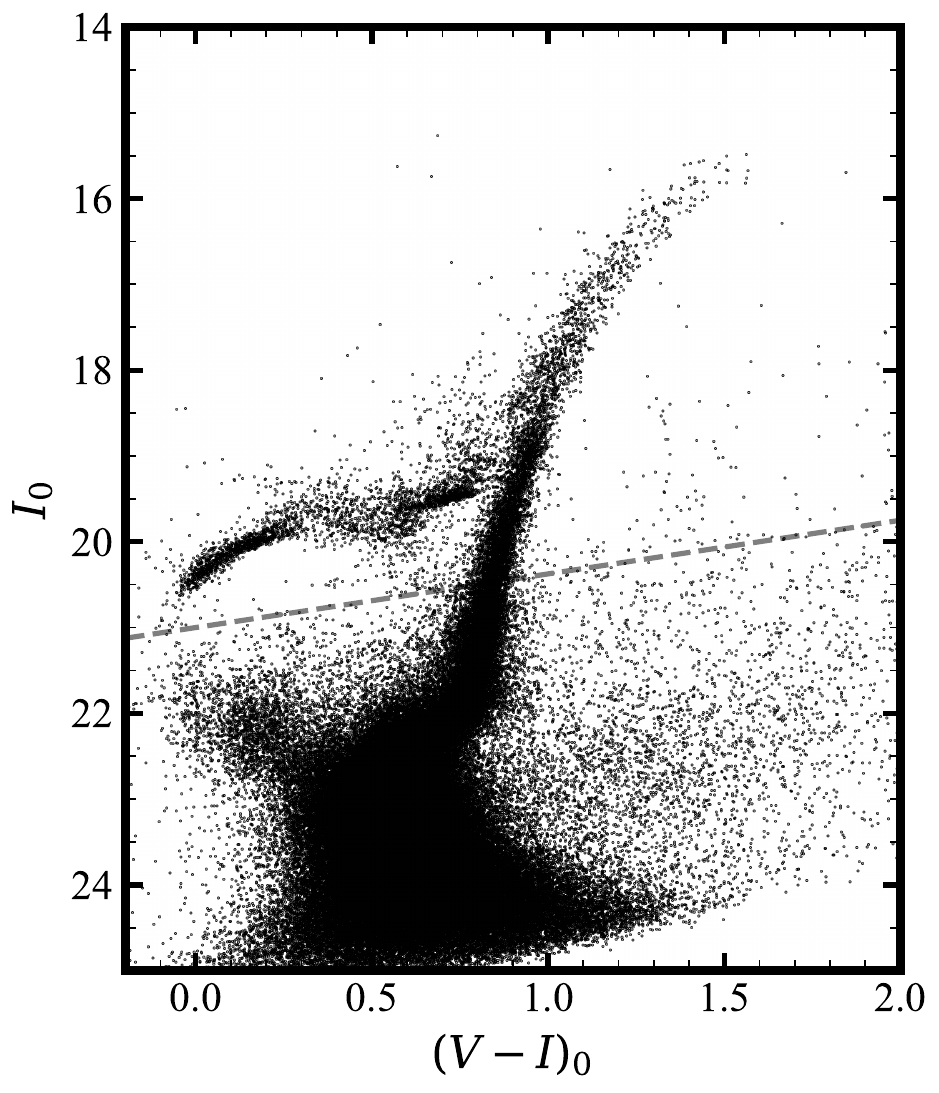}
	\caption{Sculptor Dwarf Spheroidal \textit{VI} extinction- and reddening-corrected CMD using photometry from Magellan-Baade IMACS imaging. Foreground sources are removed using proper motions from Gaia EDR3 with its depth plotted as a black dashed line (see \autoref{sec:appen_a} for details). IMACS instrumental magnitudes are calibrated to the Stetson-Landolt \textit{VI} system as described in \autoref{sec:Photo}.}
	\label{fig:cmd}
\end{figure*}

Lastly, sources confirmed to be foreground from their Gaia \citep{Gaia_2016} EDR3 \citep{Gaia_2021} proper motions \citep{Lindegren_gaia_2021, Fabricius_2021, Torra_2021} have been removed from the catalog (see \autoref{sec:appen_a} for details).
The calibrated, reddened-corrected, and foreground-cleaned composite CMD for the Sculptor dwarf spheroidal galaxy is presented in \autoref{fig:cmd}.

\section{\label{sec:trgb} \bf TRGB Distance Determination}

Low-mass stars evolving from the main-sequence onto the RGB have an isothermal helium core supported by electron degeneracy. Over time, this core increases in mass until the temperature rises high enough for helium to ignite in a brief thermonuclear runaway event. The tremendous energy output from this event lifts the electron degeneracy and the star rapidly evolves off the RGB onto the HB or Red Clump. This transition causes a sharp discontinuity in the LF for the RGB. This discontinuity is observationally identified as the TRGB \citep{Lee1993, Salaris2005}.

\subsection{TRGB Edge Detection and Measurement \label{sec:trgb_dist_measure}}

The TRGB edge detection follows the procedure described by \citet{Hatt2017}, who measured the TRGB for IC 1613. First, the Sculptor $I$--band LF was binned at 0.01 mag intervals. Then, using the GLOESS smoothing filter \citep[Gaussian-windowed, Locally Weighted Scatterplot Smoothing; first introduced by B.F.M. in][]{Persson2004}, the binned LF was smoothed in order to suppress Poisson noise. GLOESS smoothing is a non-parametric, model-independent method, which uses all available data points (suitably weighted) to estimate the values of a discrete, non-parametric, array smoothly describing those data. GLOESS is robust in measuring the TRGB due to the non-localized treatment of bins in the LF. Finally, a $\{-1, 0, +1\}$ Sobel edge-detection kernel, weighted by the Poisson signal-to-noise in each bin, was convolved with the smoothed LF, approximating a measurement of the first derivative. The place of greatest change, identified by the filter response, corresponds to the position of the TRGB \citep{Lee1993}.

In contrast to \citet{Hatt2017}, our TRGB detection does not require injection-recovery tests of artificial star LFs. Firstly, the potential for systematics in the photometry is negligible. At the TRGB magnitude, the photometric errors are on the order of 0.01 mag. Additionally, the photometric depth typically reaches $V = 22$ mag and $I = 23$ mag, which is several magnitudes below/fainter than the TRGB magnitude, ensuring that the photometry is complete at the TRGB.

Furthermore, the possibility for astrophysical systematic errors is minimal. The muted star formation history of the Sculptor dwarf spheroidal \citep{deBoer2011, Bettinelli2019} indicates that the contribution from young and intermediate-age stellar populations to the color-magnitude diagram, and thus our TRGB measurement, will be negligible. This is confirmed by inspection of \autoref{fig:cmd}, where the HB and RGB features are the most densely-populated features above the main sequence. We further note that there is an observable scatter in Sculptor's RGB. This dispersion along the RGB is an expected result of Sculptor’s star formation history, astrophysical in nature, and not a result of photometric scatter, as has been noted in previous studies \citep{Bettinelli2019}.

\begin{figure*}[ht!]%
\onecolumngrid
	\centerline{\includegraphics[width=1.0\textwidth]{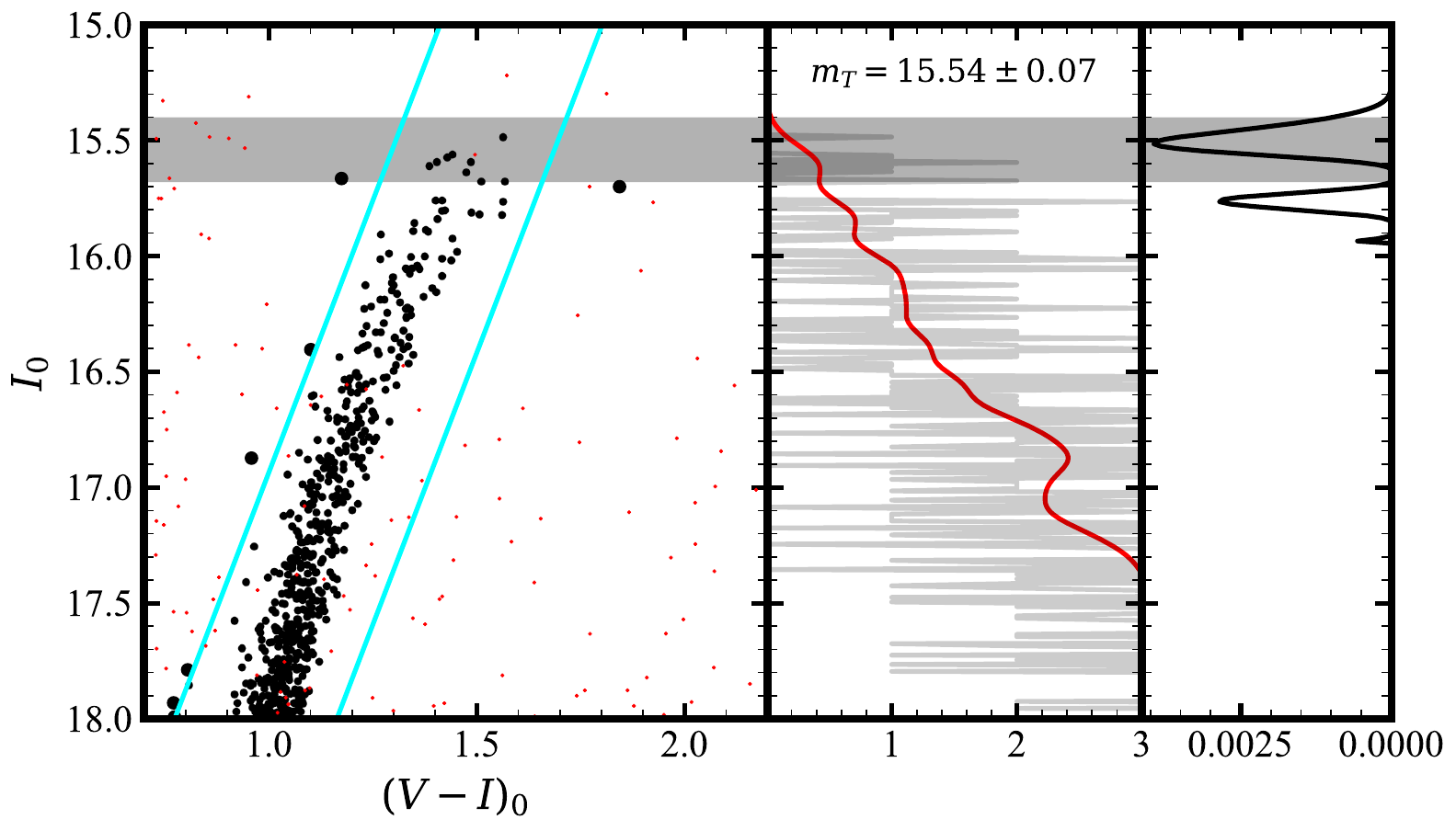}}
	\caption{The detection of the TRGB in Sculptor. \textbf{Left:} The reddening-corrected CMD of Upper-RGB stars ($I_0 < 18.0$~mag) in our IMACS catalog. The cyan lines define the color-magnitude selection used to isolate RGB stars for the TRGB detection. Red dots are foreground sources removed with Gaia EDR3 proper motions. Open black points are member sources excluded by the color cut. \textbf{Center:} The GLOESS-smoothed LF (red, solid) with a smoothing parameter of $\sigma_s = 0.1$ and 0.01 mag binned LF (grey). \textbf{Right:} Edge response function. The measured TRGB is $I_0 = 15.54 \pm 0.07_\mathrm{stat} \pm 0.02_\mathrm{sys}$~mag.}%
	\label{fig:sculptor_trgb}%
\end{figure*}

\autoref{fig:sculptor_trgb} shows the reddening-corrected measurement of the TRGB using the GLOESS smoothing filter and Sobel edge-response function. The CMD is shown in the left panel, with the shaded blue regions representing the adopted boundaries of the RGB LF. This region was manually selected to exclude the few non-RGB stars in this region of the CMD, since a small number of contaminating sources could, in principle, influence the measured tip magnitude. An empirical slope of $m_\mathrm{RGB} = -4.7$ mag color$^{-1}$ is estimated, and used to define the bounds of the region selected to reject a very small number of non-RGB stars (open black circles in \autoref{fig:sculptor_trgb}). In panel (c), the output edge detection response function is displayed. The TRGB magnitude was found to monotonically decrease with increasing smoothing kernel width, suggesting an oversmoothing bias for large ($\sigma>0.08$~mag) kernel widths. Thus, we adopt the appropriately smoothed TRGB locus determined with a kernel width $\sigma=0.05$~mag, $I_0^\mathrm{TRGB} = 15.54$~mag. The entirety of this dependence on adopted kernel width is folded into the error budget.

We perform a number of tests to estimate the full statistical uncertainty budget for our adopted detection associated with our TRGB measurement methodology. These tests include varying our color-magnitude selection, cuts on galactocentric radius, the edge measurement with or without the Poisson weighting scheme, and smoothing kernel width. The full range of TRGB magnitudes from the tests was 0.06~mag. We jack-knife the TRGB stars within 0.4 mag of the adopted TRGB magnitude. The full range of jack-knifed realizations of the TRGB magnitude is 0.03~mag. Combined with the edge detection tests, we adopt a 0.07 statistical uncertainty on our TRGB magnitude determination.

The systematic uncertainty on the TRGB magnitude determination is adopted as half of the foreground extinction in the $I$--band, rounded up (0.02~mag). Combined, we adopt a final $I_0^\mathrm{TRGB} = 15.54$~mag $\pm 0.07_\mathrm{stat} \pm 0.02_\mathrm{sys} $ mag.

We estimate the distance to Sculptor by adopting a TRGB zero-point calibration equal to $M_{I}^{\mathrm{TRGB}} = -4.047 \pm 0.022_\mathrm{stat} \pm 0.039_\mathrm{sys}$ mag \citep{Freedman2020}. This calibration is consistent with values of the TRGB zero-point from other independent studies. \citet{Lee1993} adopted $M_{I}^{TRGB} \approx -4$ mag. \citet{Bellazzini2001} calculated an $M_{I}^{TRGB} \approx -4.04 \pm 0.12$ mag for an [Fe/H] $\sim$ -1.7 dex. Similarly, \citet{Rizzi2007} found $M_{I}^{TRGB} = -4.050$ mag for Sculptor using an adopted [Fe/H] $= -1.74$.

\citet{Freedman2020} include a term to consider contributions caused by differing photometric zero-points. We similarly adopt an additional 0.02 mag systematic effect for potential differences between the Stetson-Landolt and the OGLE calibrators. We further adopt a 0.01 mag uncertainty for any potential extinction law variations in our adopted zero-point calibration. We find a true TRGB distance modulus to Sculptor of $\mu_0^\mathrm{TRGB} = 19.59 \pm 0.07_\mathrm{stat} \pm 0.05_\mathrm{sys}$ mag. We summarize the contributions to the statistical and systematic error budget in \autoref{sec:error_summary}.

\subsection{Metallicity Effects}

TRGB measurements have traditionally been made in the $I$--band for bluer ($V-I < 2$~mag) TRGB stars, where the TRGB dependence on metal content is, in practice, flat \citep{DaCosta1990}. For in-depth explorations of the $I$--band TRGB's color dependence, see \citet{Rizzi2007}, \citet{Madore2009}, \citet{Jang2017}, and \citet{Gorski_2018}. We note that Sculptor is metal-poor. \citet{Clementini2005} found an average metallicity of $\mathrm{[Fe/H]} = -1.83$ dex based on their sample of RRL stars using the calibration equations from \citet{Zinn1984}. \citet{Kirby2009} found an average metallicity of $\mathrm{[Fe/H]} = -1.58$ dex from abundance measurements of nearly 400 medium-resolution spectra of RGB stars. Recently, high-resolution abundances of 99 RGB stars from \citet{Hill2019} indicate a mean metallicity of $\mathrm{[Fe/H]} = -1.57$ dex. The luminosities of metal-rich TRGB stars are strongly coupled to metallicity, but this dependence is weakened in metal-poor populations, specifically in the $I$--band \citep{Serenelli2017}. Thus, any residual metallicity dependence that could bias the $I$--band result is minimized.

To test this, we consider a quadratic color-based correction to the TRGB magnitude using ground-based $V$ and $I$ magnitudes \citep{Jang2017}. \citet{Hoyt2021} showed that this quadratic TRGB color calibration provides the best fit to the TRGB magnitude-color relation as observed in Local Group dwarfs from the ground. Based on the $\Delta (V-I) = 0.3$~mag between the mean color of the TRGB stars that define the adopted \citet{Freedman2020} zero-point and the mean TRGB color measured in Sculptor, the color-correction as predicted by the \citet{Jang2017} calibration amounts to $\delta M_I = 0.010$~mag. This effect is well below the uncertainties of our TRGB measurement, thus confirming that a color correction is not required.

\section{\label{sec:calibrate} \textbf{RR Lyrae}}

\subsection{Light Curves and Mean Magnitudes}
\subsubsection{\label{RRL Samples} Sample Selection}
RRLs are identified in our IMACS catalog by referencing to the RA/DEC coordinates of RRLs in the extensive and high-quality sample presented in \citet[][hereafter MV15]{Martinez-Vazquez2015}. MV15 created a photometric database on the $BVI$ Johnson/Cousins photometry system from images spanning $\sim$24 years. In their analysis, they identified 536 RRLs, divided into 289 RRab, 197 RRc, and 50 RRd variables. Their penultimate sample size is 520 RRLs, having rejected 16 outliers which fall more than 2.5$\sigma$ from the $V$ magnitude average of the total population (20.13 mag). For their analysis, MV15 adopt a ``clean'' sample, which removes all RRL that have noisy light curves and/or low-quality phase coverage. This further reduces their sample to 167 RRab and 123 RRc for a total of 276 RRLs.

\begin{figure*}
\centering
	\includegraphics[width=1\textwidth]{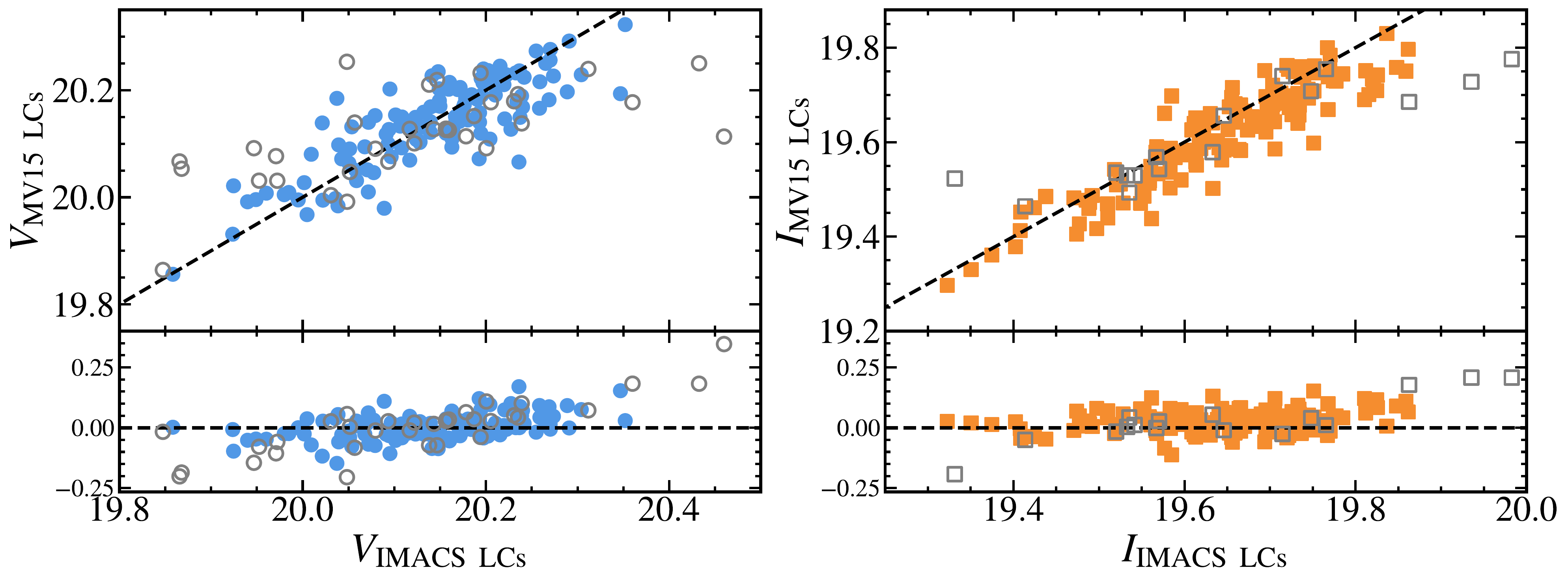}
    \caption{Comparison of mean magnitudes determined from literature MV15 to new IMACS light curves in the $V$--band (blue circles, left) and $I$--band (orange squares, right). Sources that were observed for only one epoch with IMACS were removed and a further 3$\sigma$ clip is applied. Clipped RRLs are denoted with an open grey circle. Residuals are shown in the bottom panel and the average and standard deviation of the offsets are $\Delta V = 0.003 \pm 0.053$ and $\Delta I = 0.03 \pm 0.04$ mag.}
    \label{fig:imacs_mv15_comp}
\end{figure*}

In this work, we begin by cross-matching the 276 RRab and RRc from the ``clean" MV15 sample. 261 RRLs, 155 RRab and 106 RRc are found in the IMACS field of view. RRLs more than 2.5$\sigma$ from the population mean Wesenheit $W_{I, I-V}$ magnitude (18.89 mag) are also removed. This produces a sample size of 260 RRLs, split between 154 RRab and 106 RRc stars. A final selection cut is applied to our sample, informed by our template fitting procedure (\autoref{sec:template_comp} and \autoref{fig:imacs_mv15_comp}). This leaves a final sample of 118 RRab stars to use for a distance determination.

RRc PLRs have intrinsically higher scatter than that exhibited by the RRab PLRs. For this reason, many previous studies have typically only reported final estimates based on a fundamentalized (FU+FO; RRc+RRab) population, where the periods of first overtone (FO, RRc) pulsators are shifted to fundamental (FU, RRab) RRL periods \citep[e.g.,][]{Braga2015}. However, this added dispersion from the RRc variables is undesirable, and RRab PLRs with a sufficient population, as in this study, work equally well for distance measurements so we choose to use only the RRab stars in subsequent analysis.

\subsubsection{\label{LC fits} Light-Curve Fitting}

For our sample, the typical number of phase points observed as part of the IMACS program was $\bar{N}_\mathrm{epoch} = 3 \pm 1$, where contiguous observations, e.g., typically back-to-back exposures taken within 10 min of each other are binned into a single phase point. Owing to our sparse sampling rate, we adopt periods and $V$--band amplitudes from MV15, which were determined from well sampled $BV$ photometry (83 and 52 average phase points, respectively).

With the light curve parameters from MV15 in hand, we then proceed to fit template light curves to our IMACS phase-points to determine template mean magnitudes. These templates are based on a formalism described in \citet{Freedman1988} and \citet{Freedman2010}. As part of the Three-hundred MilliMeter Telescope project \citep[TMMT;][]{Beaton2016,Monson2017}, high-cadence, high-precision optical (covering $BVRIJHK$ and Spitzer 3.6 $\mu$m and 4.5 $\mu$m) light curves of 55 RR Lyrae were acquired and combined with extensive literature observations \citep{Monson2017}. These light curves were combined and the GLOESS technique was used to produce smooth light curves with small-scale features intact \citep{Monson2017}. \citet{Beaton2016} demonstrated how the template technique formalism can generate predictive templates for non-Blashko RRab and RRc type variables in ten photometric bands. These templates can be used to derive mean optical intensities from sparsely sampled light curves.

\begin{figure}
\centering
	\includegraphics[width=0.475\textwidth]{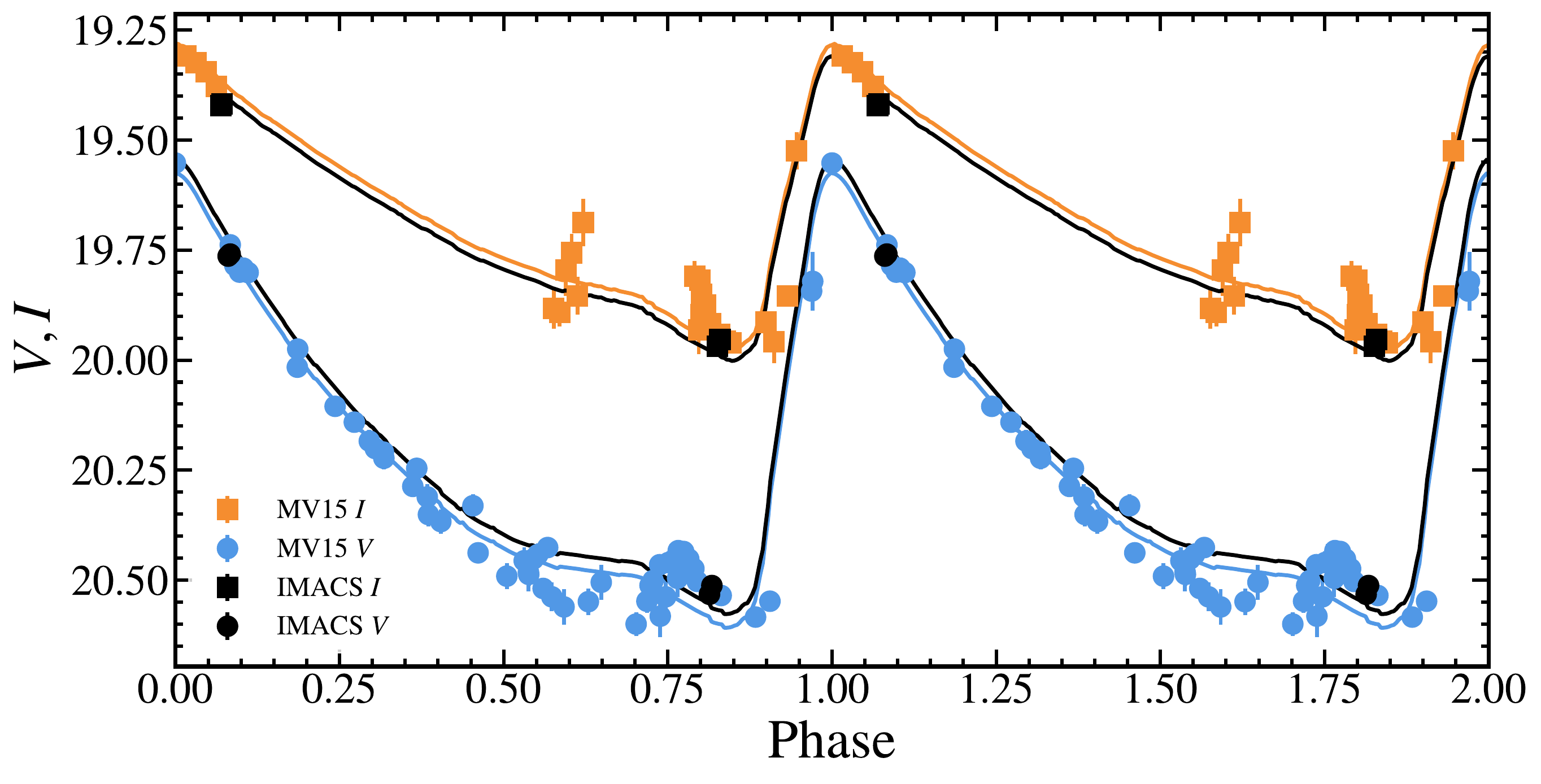}
    \caption{Template light curve fitting. For the RRab scl-CEMV144, $VI$ light curves are plotted as squares and circles, respectively, with literature MV15 photometry (blue, orange for $V$, $I$, respectively) and corresponding IMACS photometry (black). The solid lines are TMMT template fits to the MV15 (colored) or IMACS (black) light curves.}
    \label{fig:templates}
\end{figure}

Next, we transformed the templates from normalized units to match the observed photometric data. $V$--band amplitudes, $a_\mathrm{V}$, were adopted from the well-sampled MV15 light curves. To then compute the $I$--band amplitudes, $a_\mathrm{I}$, for each RRL, an amplitude ratio $a_\mathrm{I}/a_\mathrm{V} = 0.65 \pm 0.01$ was adopted (Beaton et al., in prep). The normalized templates were multiplied by the corresponding $a_\mathrm{V}$ and $a_\mathrm{I}$ amplitudes, respectively, to begin the minimization process.

The templates were then calibrated to the observed light curve by fitting for a magnitude offset. A template value for each phase point in the observed light curve was calculated by interpolating between the closest two template points. The total magnitude offset is the average of the difference between interpolated template and observed light curve at each phase point. 

We shifted the template onto the observed magnitude scale using this offset and calculated a flux-weighted average magnitude for each RRL from the best-fit template light curve.

\subsubsection{Comparison with MV15 Magnitudes \label{sec:template_comp}}

In the \autoref{sec:appen_b}, we redetermined mean magnitudes from MV15 light curves with the TMMT templates and compare to the mean magnitudes originally quoted in MV15. Very good agreement is established with $\Delta V = 0.002 \pm 0.019$ and $\Delta I = 0.002 \pm 0.031$ mag, where the ``uncertainties'' quoted are the standard deviations and not the standard errors. There is also no trend observed in the residuals (\autoref{fig:template_compare}).

With consistency established in the template-fitting process, we then fit the IMACS light curves and determined their mean magnitudes determined with the TMMT templates. This  produced an entirely independent set of mean magnitudes based on MV15's determined periods, on the \emph{same photometric system as the other distances presented here}. This is critical to do---as opposed to directly adopting the MV15 magnitudes or light curves---because differences in photometric zero-points can exceed $0.02-0.03$ mag across various definitions of the J-C photometric system deployed at different sites, telescopes, and instruments. Thus, our approach of using only IMACS phase points will reduce systematic uncertainties incurred by mixing heterogeneous photometry, with the trade-off being a concomitantly larger scatter than the MV15 PL relations, which are based on better sampled light curves.

In \autoref{fig:imacs_mv15_comp}, a comparison between the mean magnitudes of the MV15 light curves  and the mean magnitudes of the IMACS light curves, both found using the TMMT templates, is shown. Here we found a small systematic offset and applied another selection cut prior to the final PL/PW and PLZ/PWZ fits. All RRab stars with only one epoch were removed. 3$\sigma$ outliers from this comparison are clipped, leaving a final, optimal sample of 118 RRab stars to determine a distance.

\subsection{\label{PL/PW R} RRL I-band and Wesenheit Period Metallicity Relations}

Wesenheit magnitudes were calculated for each RRL using the original, reddened magnitudes ($V$ and $I$). Wesenheit magnitudes are defined to be reddening free---under the assumption of a universal reddening law---nominally reducing reddening uncertainties to zero \citep{Madore1982, Braga2015, Marconi2015}. The Wesenheit magnitude is defined in \citet{Madore1982} as:
\begin{equation} \label{eq:wesenheit}
W_{M_x,M_y-M_x} = M_x + \xi(M_x - M_y)
\end{equation}
where $\xi$ is a coefficient that describes the color absorption and excess ratio between $M_x$ and $M_y$. For W$_{I, V - I}$, the color coefficient from \citet{Neeley2019}'s Table 3, 1.467, is adopted.

We also adopted from \citet{Neeley2019} a PLZ/PWZ relation of the form
\begin{equation}
m = a + b (\text{log} P + 0.3) + c(\mathrm{[Fe/H]} + 1.36)
\end{equation}
and fit for the $a$ while adopting $b$ and $c$ directly from \citet{Neeley2019}'s Table 3. The PL/PW relations are of the same form without the metallicity term. For the PLZ/PWZ relations, we adopt a mean metallicity of $\mathrm{[Fe/H]} = -1.83$ dex, taken from \citet{Clementini2005}, which is on the \citet{Zinn1984} scale. The relations from \citet{Neeley2019} are also on the same scale.

\begin{figure*}[ht!]%
	\centerline{\includegraphics[width=1\textwidth]{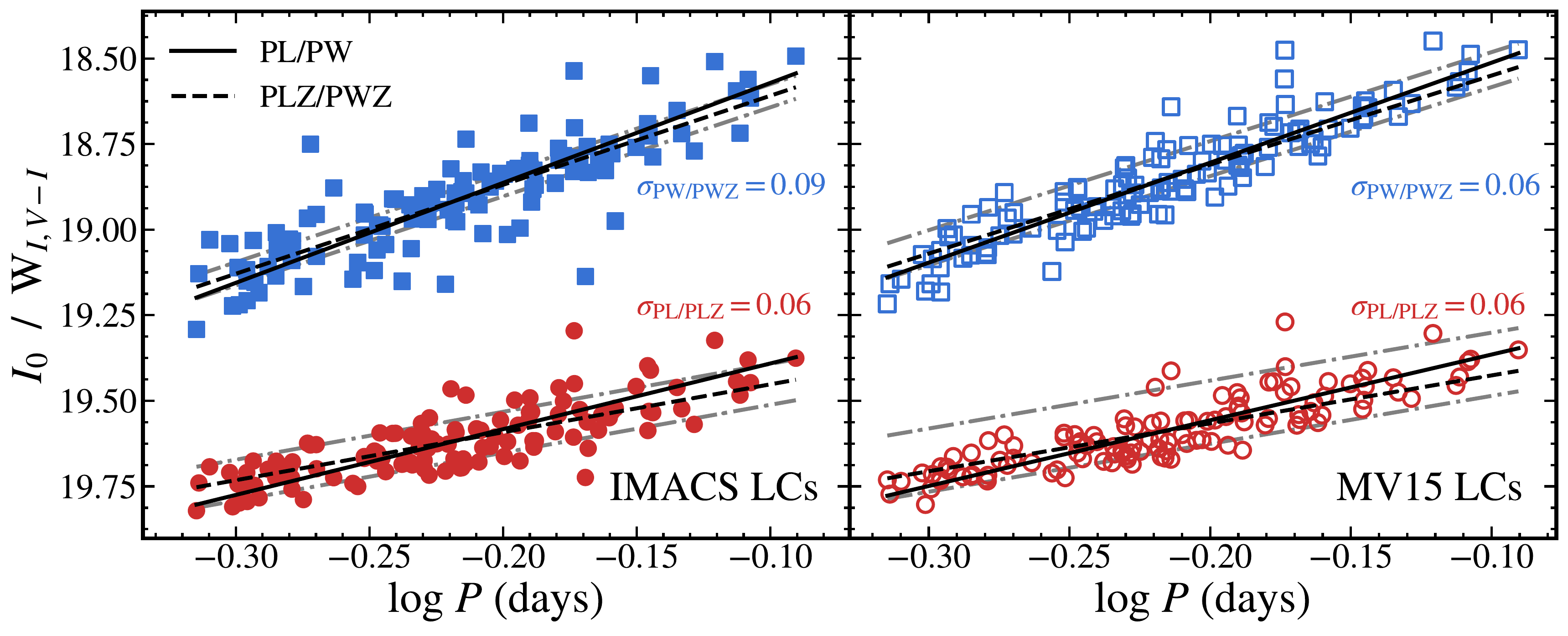}}
	\caption{RRab period-luminosity relations based on the new IMACS light curves (left) vs. literature LCs (right) from \citet[][MV15]{Martinez-Vazquez2015}. Both sets of mean magnitudes were determined using TMMT template fits to the LCs. PL relations are shown in $I_0$ (red circles) and $W_{I, V-I}$ (blue squares). Solid lines are best fit PL/PW relations and dotted lines are the best fit PLZ/PWZ relations \citep[slopes and metallicity terms adopted from][]{Neeley2019}. Dotted gray lines are shown to represent the full range of metallicity considered.} The MV15 light curve $I$--band PL relation (right) exhibits smaller scatter than the IMACS equivalent (left) due to the higher sampling rate in their LCs. The results when using either the new IMACS LCs or the MV15 LCs are in agreement with each other (see \autoref{PL/PW R}).%
	\label{fig:PL_relations}%
\end{figure*}

The adopted mean metallicity is broadly consistent with expectations from the literature. \citet{Clementini2005} found an RMS scatter of $\sigma = 0.26$ dex, for a metallicity range of $-2.09$ to $-1.57$ dex in their RRL sample. We conservatively adopt this range as potential metallicities to be considered. The upper limit on this range is consistent with the theoretical upper limit expected from RGB stars. \citet{Hill2019}, using high resolution spectra of a sample of 99 RGB stars, calculates a mean of $\mathrm{[Fe/H]} = -1.57$. We consider this metallicity range (shown in \autoref{fig:PL_relations}) as a potential source of systematic uncertainty in our PLZ fits. Changing the adopted metallicity from $\mathrm{[Fe/H]} = -2.09$ to $-1.57$ dex changes the distance modulus of the PLZ relation by 0.12 mag and of the PWZ relation by 0.07 mag. We adopt half of these total shifts as the metallicity contribution to the systematic error for each fit.

Finally, we apply a foreground reddening correction (\autoref{sec:Photo}) to the $I$--band mean magnitudes for our PL/PLZ relations and adopt half of this value, rounded up, as systematic uncertainty (0.02 mag). Furthermore, we adopt a 0.01 mag uncertainty to account for potential deviations from the adopted mean Galactic extinction law ($R_V=3.1$) and an additional 0.01 mag uncertainty for extinction law variations in our adopted zero-point calibrations. The PL relations are shown in \autoref{fig:PL_relations} with the best-fit parameters tabulated in \autoref{tab:PLR_fit_params}.

The left panel of \autoref{fig:PL_relations} shows the fit to both the PL/PW and PLZ/PWZ relations for our final RRab sample, using periods and amplitudes from MV15. RRab $I_0$--band magnitudes are plotted as red circles and Wesenheit $W_{I, V-I}$ magnitudes as blue squares. The best-fit PL/PW and PLZ/PWZ relations are shown in solid black and dashed black lines, respectively. The right panel of \autoref{fig:PL_relations} is the same, except that in this case the mean magnitudes were derived from the MV15 light curves. These fits are shown as a comparison to our IMACS-only analysis. As previously noted, an increase in the scatter of the relations is seen due to the decrease in light curve sampling. The best-fit zero-points for the PL and PLZ relations between the IMACS and MV15 light curves each differ by $<$0.03 mag. The best-fit zero-points for PW and PWZ relations both differ by $<$0.06 mag, but this is well within the uncertainty of our final distance moduli measurement. Furthermore, the RMS of our best fit of the IMACS light curves to the PL/PW and PLZ/PWZ relations are 0.06/0.09 and 0.06/0.09, respectively. The RMS values for the PL/PW and PLZ/PWZ relation using the MV15 LCs are comparable at 0.06/0.06 and 0.06/0.07, respectively. Our larger Wesenheit scatter is due to the better sampling of the MV15 $V$--band light curves.

In fitting to the \citet{Neeley2019} relations, we adopt two additional sources of systematic uncertainty. We adopt in quadrature the reported errors for the \citet{Neeley2019} zero-point parameters, 0.03 mag for the PL/PLZ and 0.02 mag for the PW/PWZ relations, as additional systematics. We also adopt an uncertainty of 0.01~mag to account for potential zero-point differences between the IMACS photometry and the \citet{Neeley2019} photometry. Notably, this small value is still likely conservative because the TMMT photometry that defines the \citet{Neeley2019} calibration was calibrated to Stetson's Landolt standards, and the IMACS photometry presented here was also calibrated onto the same Stetson-Landolt system.

\begin{deluxetable}{ccccc}[!t]
    \small
    \centerwidetable
	\tablecaption{RRL PL/PW and PLZ/PWZ Fit Parameters\label{tab:PLR_fit_params}}
    \tablehead{
    \colhead{Band} & \colhead{$a$} & \colhead{$b$\tablenotemark{a}} & \colhead{$c$\tablenotemark{a}} & \colhead{[Fe/H]\tablenotemark{b}}}
    \startdata
    \multicolumn{5}{c}{\textit{PL/PW Relations}} \\
    $I$ & $19.77 \pm 0.01$ & $-1.92 \pm 0.41$ & & \\
    $W_{I,V-I}$ & $19.16 \pm 0.01$ & $-2.92 \pm 0.30$ \\
    \multicolumn{5}{c}{\textit{PLZ/PWZ Relations}} \\
    $I$ & $19.84 \pm 0.01$ & $-1.40 \pm 0.30$ & $0.23 \pm 0.04$ & $-1.83$ \\
    $W_{I,V-I}$ & $19.19 \pm 0.01$ & $-2.60 \pm 0.25$ & $0.13 \pm 0.03$ & $-1.83$ \\
    \enddata
    \tablenotetext{a}{\citet{Neeley2019}}
    \tablenotetext{b}{\citet{Clementini2005}}
    \end{deluxetable}

\subsection{\label{sec:RRL_Results} RRL Distance Moduli}
In this section, we report calculated distance moduli to Sculptor. From our PL and PW relation fits, we find distance moduli of $19.60 \pm 0.01_\mathrm{stat} \pm 0.07_\mathrm{sys}$ and $19.58 \pm 0.01_\mathrm{stat} \pm 0.05_\mathrm{sys}$ mag, respectively. Similarly, we find distance moduli of $19.67 \pm 0.01_\mathrm{stat} \pm 0.07_\mathrm{sys}$ and $19.60 \pm 0.01_\mathrm{stat} \pm 0.05_\mathrm{sys}$ mag for our PLZ and PWZ relations, respectively. Here, the statistical errors are the uncertainties on $a$, the zero-point parameter from the least-squares fit, as reported in \autoref{tab:PLR_fit_params}. These statistical uncertainties were verified with bootstrap simulations. Systematic errors are the quadrature sum of every contribution as described above. For a summary of these sources, see \autoref{tab:error_budget} in \autoref{sec:error_summary}. For our final RRL distance determination, we prefer to explicitly account for metallicity effects so we do not adopt the PL/PW distances, which are nevertheless presented for completeness. Of the metallicity-dependent relations, we find effects of metallicity on the PWZ distance is 66\% that of the effect in the $I$--band PLZ, so we adopt the PWZ distance as our nominal RRL distance to find $\mu_{W_{I,V-I}}^\mathrm{RRL} = 19.60 \pm 0.01_\mathrm{stat} \pm 0.05_\mathrm{sys}$ mag.

\section{\textbf{The Horizontal Branch} \label{sec:zahb}}

The HB distance to the Sculptor dSph is determined from a direct comparison of its blue HB with HBs observed in globular clusters to which trigonometric distances have been determined. This method relies on the fact each cluster defines an ``empirical isochrone" in color--magnitude space that can be calibrated observationally \citep[e.g,][]{Clem_2008}. We select the clusters M55 and $\omega$ Cen (NGC 5139), as both clusters are metal-poor ($\rm [Fe/H] = -1.94$ dex and $\rm [Fe/H] = -1.53$ dex, respectively; \citealt{1996AJ....112.1487H}), and have geometric detached eclipsing binary (DEB) distances available from the literature \citep{2014AcA....64...11K,2001AJ....121.3089T}. While the former cluster more closely approximates that of Sculptor's most ancient, metal-poor population, the latter cluster exhibits a greater dispersion in metallicity, and may better approximate the stellar population(s) observed in dSphs including Sculptor. Using these two clusters helps in quantifying the extent to which our HB distance could be biased due to metallicity effects on the HB's location in absolute color--magnitude space. We note that this method provides an update to that cited in \citet{Freedman2021}; the resulting distance modulus is 0.02 mag closer (1\% in distance).

We begin by selecting member stars in each of these clusters using proper motions from \textit{Gaia} EDR3 following the method described in \citet{Cerny_2021}. Then, we cross-matched this member-star catalog to the high-precision, homogeneous cluster photometry provided by \citet{2019MNRAS.485.3042S}. The cluster photometry is calibrated to the same Stetson-Landolt photometric system to which the Sculptor photometry was calibrated, mitigating the known systematic errors incurred when combining heterogeneously calibrated photometry. Next, we construct a horizontal branch fiducial ridgeline for each of the two clusters by sampling the (3$\sigma$-clipped) median $I-$band magnitude of the blue HB in color intervals of 0.025 mag, with the fiducial HB line computed from a spline interpolation. To absolutely calibrate the resulting fiducial HB curves, we adopt DEB distances and reddening values from the literature, $\mu_{0} = 13.678 \pm 0.1$ mag and $E(B-V) = 0.12$ for $\omega$ Cen \citep{2001AJ....121.3089T} and $\mu_0 = 13.58 \pm 0.05$ mag and $E(B-V) = 0.115$ \citep{2014AcA....64...11K} for M55.

\begin{figure*}[t!]%
	\centerline{\includegraphics[width=1\textwidth]{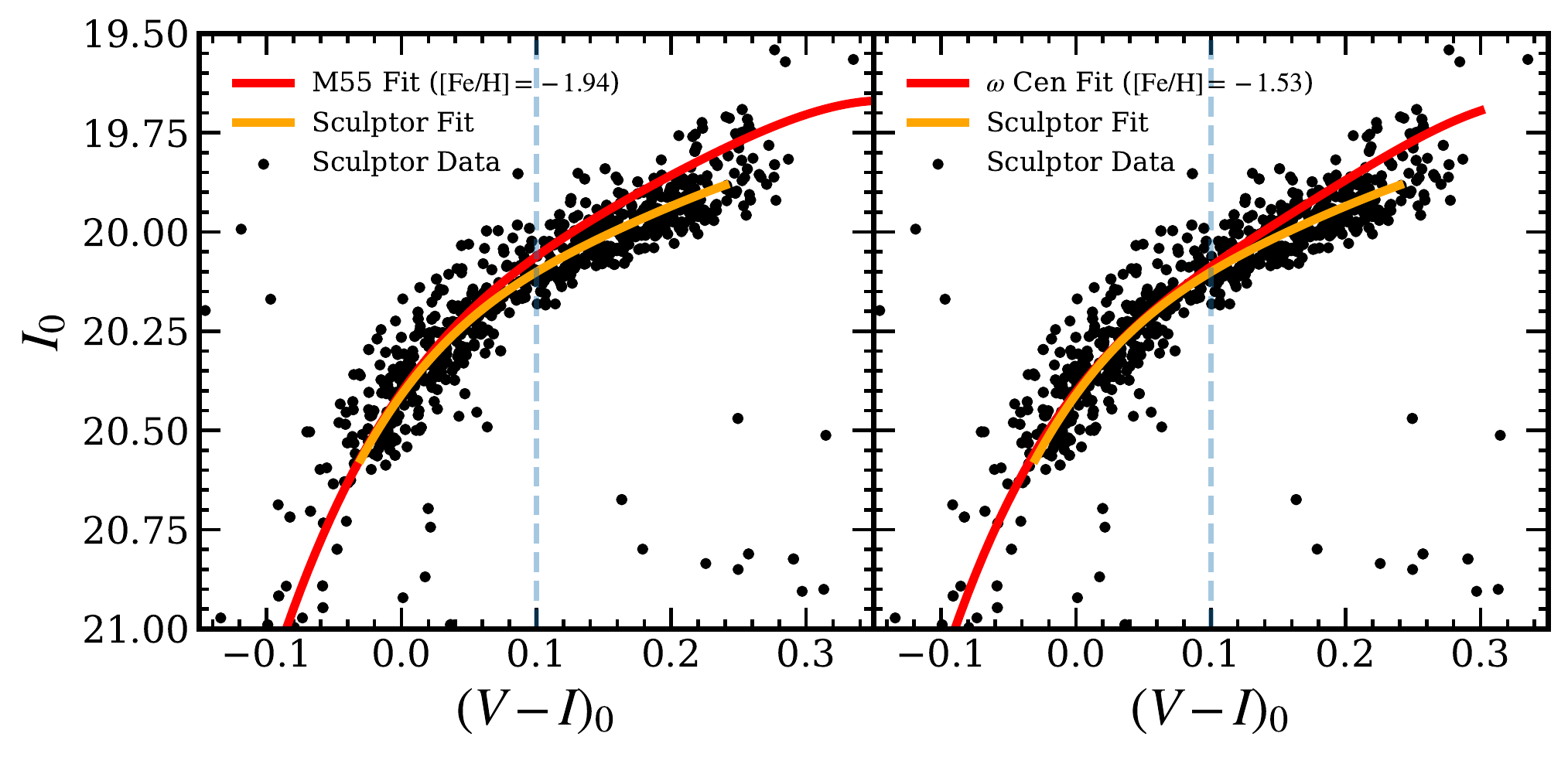}}
	\caption{Fiducial ridgeline of the BHB for Sculptor compared to that of M55 (left) and $\omega$ Cen (right). All fiducials are created on the BHB using the same process (see \autoref{sec:zahb}). The lines are the fiducials for the globular clusters (red) and for Sculptor (orange). The dashed blue line represents the upper limit of the color selection $(V - I)_0 < 0.1$ mag, over which each cluster BHB is differenced with Sculptor's to determine a distance. Final distances are derived over this blue selection because the residuals are smaller when compared to a minimization to the full BHB, which is expected given the blue BHB's predicted weaker sensitivity to metallicity and age effects. The distance modulus to Sculptor derived from M55 is $\mu_0 = 19.55$ mag and $\mu_0 = 19.49$ mag for $\omega$ Cen.}
	\label{fig:HB_fit}%
\end{figure*}

The cluster HB-fitting process is repeated for Sculptor which also has a very well-defined blue HB (\autoref{fig:HB_fit}). The fiducial cluster HBs (red curves in both panels of \autoref{fig:HB_fit}) are then fit to Sculptor's fiducial HB (orange curves in both panels of \autoref{fig:HB_fit}) via least-squares minimization. We find that the distance modulus is mildly sensitive to the color range over which the differential HB comparison is made. This is not unexpected, given the known irregular bipartite segmentation of Sculptor's HB \citep{Majewski_1999}. Thus, for both clusters, we explore the magnitude of this effect and update the error budget to account for the findings.

Using the M55 fiducial, we find a best-fit distance modulus with no restrictions on photometric color, $\mu_{0} = 19.57$ mag. If the color range is instead restricted to the bluest half of the HB, $(V-I)_{0} < 0.1$ mag (indicated by the blue dashed line in \autoref{fig:HB_fit}), we find the best-fit distance modulus to be $\mu_{0} = 19.55$ mag. Reversing the color selection to $(V-I)_{0} > 0.1$ results in $\mu_{0} = 19.60$. 
All of these results are consistent with the distances derived via the TRGB and RR Lyrae. However, we argue that the bluest result ($\mu_{0} = 19.55$ mag) is likely more accurate for two main reasons. Firstly, the residuals (after normalizing by the number of datapoints) are smallest suggesting a better match of HB populations. Secondly, it is theoretically expected that the HB is least sensitive to the metallicity and age dependent effects at bluer colors \citep[e.g.,][]{1999AJ....118.1738F} and thus the bluer selection (where the morphology is most similar) is more consistent with this prediction. Therefore, we adopt the $\mu_{0} = 19.55$ mag for the M55 distance.

For $\omega$ Cen, we repeat a similar analysis and find $\mu_{0} = 19.51$ mag when using the full color range of the HB and $\mu_{0}= 19.49$ mag when the blue color restriction of $(V-I)_{0} < 0.1$ is again applied (right panel of \autoref{fig:HB_fit}). The residuals were found to be marginally smaller for the latter case, and thus, we adopt it as our HB distance based on $\omega$ Cen. The difference between the two color selections amounts to 0.05~mag for both cluster-Sculptor distances, and we adopt that as a systematic error.

A combined HB distance of $\mu_{0} = 19.54$ mag is computed from the weighted average of the (color-restricted) individual HB distances (with weights set according to the DEB distance uncertainties). For the statistical uncertainty on this two-cluster determination, we take half the difference between our two preferred distance modulus values, $(19.55 - 19.49)/2 = 0.03$ mag. We expect this to account for any metallicity dependence of the HB location, which is consistent with prior findings over this metallicity range \citep{Dotter2010, Federici_2012}.

We discuss additional contributions to our error budget here. We first adopt a systematic zero-point uncertainty of 0.05 mag, equivalent to the total uncertainty in the more precise of the two cluster distances. Assuming systematics are shared between the two, the cumulative M55 uncertainty places an upper limit on the systematic floor of the DEB distances, thereby providing a conservative estimate of the total systematic uncertainty. We further add (in quadrature) a 0.01 mag uncertainty to account for any residual variation in the definition of this photometric system, noting again that this difference is a conservative estimate as both the cluster and Sculptor photometric catalogs are are calibrated to the same Stetson-Landolt system. We also adopt a $0.05/\sqrt{2} = 0.04$ mag uncertainty as an estimate of the intrinsic HB alignment dependence on metallicity based on the scatter of the HB-metallicity relation \citep{Dotter2010, Federici_2012}. In order to account for potential deviations from the adopted mean Galactic extinction law ($R_V=3.1$) in our measurements, we adopt a 0.01 mag uncertainty. Additionally, we further adopt a 0.01 mag uncertainty for potential extinction law variations in our adopted calibration. Finally, we adopt $15\%$ of the foreground reddening \citep{Schlegel1998, Schlafly2011} as the systematic uncertainty (0.02~mag) for the reddening to the calibrating clusters and an additional 0.02 mag uncertainty associated with foreground extinction for Sculptor. \autoref{tab:error_budget} details these sources of error.

In summary, we find $\mu_{0}^{\rm HB} = 19.54 \pm 0.03_\mathrm{stat} \pm 0.09_\mathrm{sys}$ mag. This value is in good agreement with the values derived from the TRGB and RRL PWZ relation. That these independent distance determinations are consistent suggests that the systematic uncertainties in any of the three determinations are most probably less than 0.10 mag. In the future, more accurate parallax measurements from \textit{Gaia} will improve our understanding of systematics in HB distance determination.

\clearpage

\section{\label{sec:compare} \textbf{Discussion}}

\begin{figure}
    \centering
    \includegraphics[width=0.9 \columnwidth]{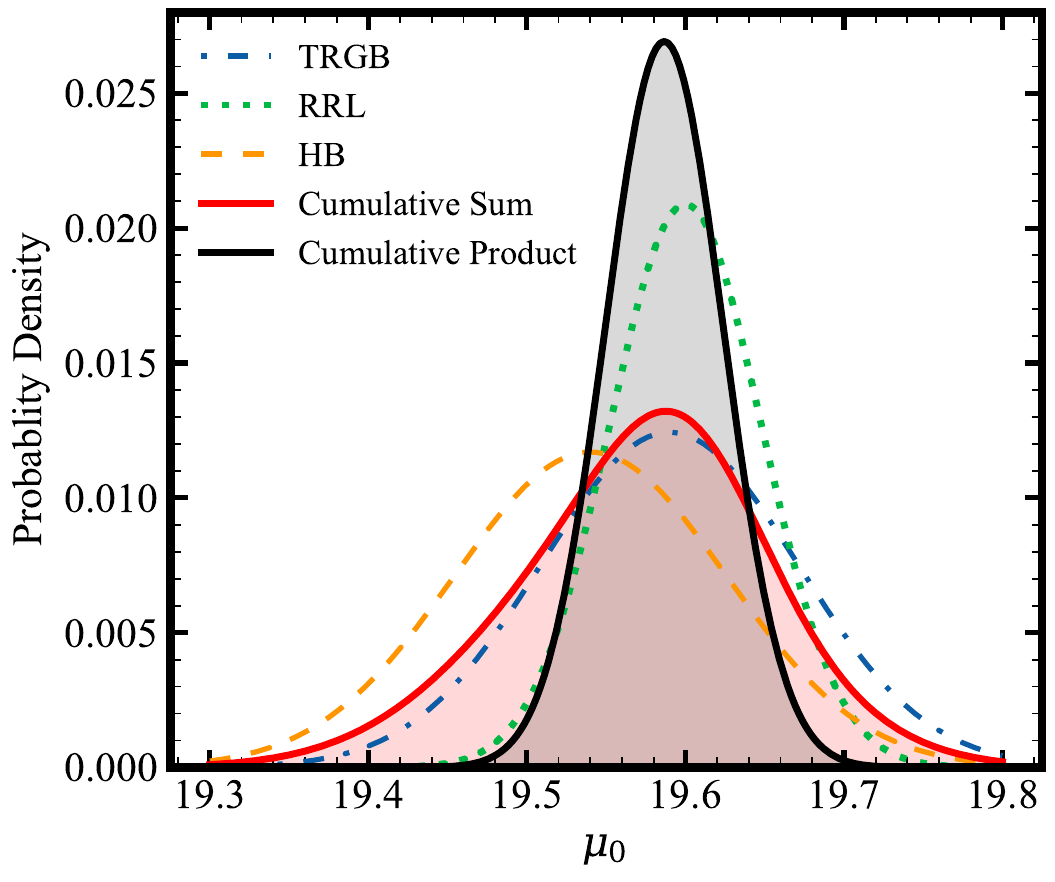}
    \caption{Overview of TRGB (blue dash-dotted), RRL PWZ (green dotted), and HB (orange dashed) distances. Also shown are two approaches to potentially combining the measurements: a product distribution (black curve) and sum distribution (red curve).}
    \label{fig:dist_dists}
\end{figure}

In this section, we provide a summary of our statistical and systematic error budgets for each distance determination and their source. The final distances and their errors are shown in \autoref{fig:dist_dists}.  We also compare our calculated distance moduli with values in the literature (\autoref{tab:compare}).

\subsection{\label{sec:error_summary} Statistical and Systematic Error Budget}

\autoref{tab:error_budget} summarizes the different sources of statistical and systematic uncertainties and reports the final error budget for each distance determination method.

\begin{deluxetable*}{cccc}[!ht]
    \centerwidetable
	\tablecaption{Summary of Distance Determination Uncertainties\label{tab:error_budget}}
    \tablehead{
    \colhead{Distance Determination Method} & \colhead{Uncertainty (mag)} & \colhead{Type\tablenotemark{a}} & \colhead{Source}}
    \startdata
    Tip of the Red Giant Branch & 0.06 & Statistical & Edge detection\tablenotemark{b} \\
    & 0.03 & Statistical & Jack-knife resampling  \\
    & 0.022 & Statistical & \citet{Freedman2020}\tablenotemark{c} \\
    & 0.02 & Systematic & Photometric ZP differences \\
    & & & between OGLE and Stetson-Landolt \\
    & 0.01 & Systematic & Extinction law uncertainty \\
    & & & in internal Sculptor reddening \\
    & 0.01 & Systematic & Extinction law uncertainty \\
    & & & in zero-point calibration \\
    & 0.02 & Systematic & Sculptor Line-of-sight Extinction \\
    & 0.039 & Systematic & \citet{Freedman2020}\tablenotemark{c} \\
    \textbf{Total Adopted TRGB ($\mu_0^\mathrm{TRGB}$)} & \textbf{0.07 and 0.05} & \textbf{Stat. and Syst.} & \\
    \hline
    RRL & 0.01 & Statistical & Bootstrap simulations \\
    & 0.01 & Systematic & Photometric ZP differences between \\
    & & & \citet{Neeley2019} and IMACS\tablenotemark{d} \\
    & 0.01 & Systematic & Extinction law uncertainty \\
    & & & in internal Sculptor reddening \\
    & 0.01 & Systematic & Extinction law uncertainty \\
    & & & in zero-point calibration \\
    & (0.02)\tablenotemark{e} & Syst. (PL/PLZ only) & Sculptor Line-of-sight Extinction \\
    & (0.06) & Syst. (PLZ only) & Metallicity effects \\
    & 0.04 & Syst. (PWZ only) & Metallicity effects \\
    & (0.03) & Syst. (PL/PLZ only) & Gaia Calibration \citep{Neeley2019} \\
    & 0.02 & Syst. (PW/PWZ only) & Gaia Calibration \citep{Neeley2019} \\
    \textbf{Total Adopted RRL ($\mu_{W_{I,V-I}}^\mathrm{RRL}$)\tablenotemark{e}} & \textbf{0.01 and 0.05} & \textbf{Stat. and Syst.} & \\
    \hline
    HB & 0.03 & Statistical & Metallicity range of GC Calibrators \\
    & 0.01 & Systematic & Photometric ZP differences between \\
    & & & \citet{2019MNRAS.485.3042S} and IMACS\tablenotemark{d} \\
    & 0.02 & Systematic & GC reddening \\
    & 0.01 & Systematic & Extinction law uncertainty \\
    & & & in internal Sculptor reddening \\
    & 0.01 & Systematic & Extinction law uncertainty \\
    & & & in zero-point calibration \\
    & 0.02 & Systematic & Sculptor line-of-sight Reddening \\
    & 0.05 & Systematic & HB fit color range \\
    & 0.04 & Systematic & Intrinsic Scatter  \citep{Dotter2010} \\
    & 0.05 & Systematic & DEB Distance Uncertainty \\
    \textbf{Total Adopted HB ($\mu_0^\mathrm{HB}$)} & \textbf{0.03 and 0.09} & \textbf{Stat. and Syst.} & \\
    \enddata
    \tablenotetext{a}{Statistical (Stat.) or Systematic (Syst.) uncertainties.}
    \tablenotetext{b}{Tests to assess effects on measured TRGB due to changes in color-magnitude region selection, cuts on galactocentric radius, and edge measurement algorithm parameters.}
    \tablenotetext{c}{Uncertainties to account for photometric, geometric, metallicity, and reddening effects in absolute TRGB measurement, taken from Table 2 of \citet{Freedman2019}.}
    \tablenotetext{d}{Note that photometric differences are minimized as both measurements are calibrated to the Stetson-Landolt system.}
    \tablenotetext{e}{We adopt the PWZ distance modulus as the final RRL determination due to the reduced metallicity effect and minimal reddening uncertainty. All values in parentheses are accounted in the PL/PLZ only and not used in the total adopted RRL distance uncertainty determination.}
\end{deluxetable*}

\subsection{\label{sec:compare_trgb} Comparison of TRGB Distances}

In recent years, there have been two other studies that have calculated the distance to Sculptor using the TRGB method. \citet{Rizzi2007} reports $\mu^\mathrm{TRGB} = 19.60 \pm 0.03$ mag in the $I$--band. \citet{Gorski2011} measured a TRGB distance moduli of $\mu^\mathrm{TRGB} = 19.72 \pm 0.19$ mag and $\mu^\mathrm{TRGB} = 19.70 \pm 0.25$ mag in the $J$ and $K$ bands, respectively. Our measured, reddening-corrected TRGB distance modulus, $\mu_0^\mathrm{TRGB} = 19.59 \pm 0.07_\mathrm{stat} \pm 0.05_\mathrm{sys}$ mag, is consistent with these literature values to within the quoted uncertainties.

\subsection{\label{sec:compare_rrl} Comparison of RRL Distances}
There have been several works in recent years that have calculated distances to Sculptor using RRL variables. \citet{Pietrzynski2008} used near-infrared light curves in the $J$ and $K$ bandpasses and reported a distance modulus of $\mu_\mathrm{RRL} = 19.67 \pm 0.02_\mathrm{stat} \pm 0.12_\mathrm{sys}$ mag. MV15, where the RRL periods and amplitudes in this study come from, adopted $\mu_\mathrm{RRL} = 19.62 \pm 0.04$ mag. More recently, \citet{Garofalo2018} found $\mu_\mathrm{RRL} = 19.60 \pm 0.02_\mathrm{stat} \pm 0.09_\mathrm{sys}$ $(\sigma_\mathrm{photo} = 0.04)$ and $\mu_\mathrm{RRL} = 19.57 \pm 0.02_\mathrm{stat} \pm 0.11_\mathrm{sys}$ $(\sigma_\mathrm{photo} = 0.04)$ mag (where $\sigma_\mathrm{photo}$ is the standard deviation of their light curve residuals) using infrared light curves with two different zero-point calibrations. Our adopted RRL distance modulus, $\mu_{W_{I,V-I}}^\mathrm{RRL} = 19.60 \pm 0.01_\mathrm{stat} \pm 0.05_\mathrm{sys}$ mag agrees with all of these literature measurements to within their reported uncertainties.

\subsection{\label{sec:compare_hb} Comparison of HB Distances}

There is one other work that has calculated the distance to the Sculptor dSph using an HB-centric method. \citet{Rizzi2007} used $V$--band ground-based photometry and selected for stars within $\sim$1~mag of the HB with color $0.2 < (V - I) < 0.6$~mag and took the average magnitude. Adopting the absolute calibration from \citet{Carretta2000}, they found a distance modulus of $19.66 \pm 0.15$ mag. Our calculated HB distance, $\mu_0^{\mathrm{HB}} = 19.54 \pm 0.03_\mathrm{stat} \pm 0.09_\mathrm{sys}$ mag is consistent with this result.

\movetabledown=1.5in
\begin{table*}[ht]
\begin{rotatetable}
	\caption{Comparison with Independent Studies\label{tab:compare}}
    \begin{center}
   \begin{tabular}{l c l l c c}
    \toprule
    \small
    ~~~~Method & Filter & ~~~~~$\mu$ (mag) & Adopted $\mu$ (mag) & Zero--Point & Reference \\
    \hline
    RRab PWZ & $W_{I, V-I}$ & $19.60 \pm 0.01 \pm 0.05$ & & \citet{Neeley2019} & this study \\
    HB & $I$ & $19.54 \pm 0.03 \pm 0.09$ & & \cite{2001AJ....121.3089T, 2014AcA....64...11K} & this study \\
    TRGB & $I$ & $19.59 \pm 0.07 \pm 0.05$ & & \citet{Freedman2020} & this study \\
    \hline
    RRab PW & $W_{V, B-V}$ & $19.59 \pm 0.05$ &  & \citet{Martinez-Vazquez2015} & MV15 \\
    RRc PW & $W_{V, B-V}$ & $19.67 \pm 0.1$ &  & \citet{Martinez-Vazquez2015} & MV15 \\
    RRab+RRc PL & $W_{V, B-V}$ & $19.61 \pm 0.05$ & $19.62 \pm 0.04$ & \citet{Martinez-Vazquez2015} & MV15 \\
    RRab+RRc PL & $K$ & $19.64 \pm 0.02 \pm 0.15$ & & \citet{Sollima2006} & \citet{Pietrzynski2008} \\
    RRab+RRc PL & $K$ & $19.72 \pm 0.02 \pm 0.13$ & & \citet{Bono2003} & \citet{Pietrzynski2008} \\
    RRab+RRc PL & $K$ & $19.68 \pm 0.02 \pm 0.12$ & & \citet{Catelan2004} & \citet{Pietrzynski2008} \\
    RRab+RRc PL & $J$ & $19.66 \pm 0.02 \pm 0.12$ & $19.67 \pm 0.02 \pm 0.12$ & \citet{Catelan2004} & \citet{Pietrzynski2008} \\
    RRab PLZ & 3.6$\mu$m & $19.60 \pm 0.02 \pm 0.09$ & $19.60 \pm 0.02 \pm 0.09$ & \citet{Neeley2017} & \citet{Garofalo2018} \\
    RRab+RRc PLZ & 3.6$\mu$m& $19.68 \pm 0.02 \pm 0.11$ & & \citet{Muraveva2018} (\textit{HST}) & \citet{Garofalo2018} \\
    RRab+RRc PLZ & 3.6$\mu$m & $19.57 \pm 0.02 \pm 0.11$ & $19.57 \pm 0.02 \pm 0.11$ & \citet{Muraveva2018} (\textit{Gaia}) & \citet{Garofalo2018} \\
    HB & $V$ & $19.66 \pm 0.15$ & $19.66 \pm 0.15$ & \citet{Rizzi2007} & \citet{Rizzi2007} \\
    TRGB & $I$ & $19.61 \pm 0.04$ & $19.61 \pm 0.04$ & \citet{Rizzi2007} & \citet{Rizzi2007} \\
    TRGB & $J$ & $19.72 \pm 0.08 \pm 0.11$ & $19.72 \pm 0.08 \pm 0.11$ & \citet{Valenti2004} & \citet{Gorski2011} \\
    TRGB & $K$ & $19.70 \pm 0.08 \pm 0.17$ & $19.70 \pm 0.08 \pm 0.17$ & \citet{Valenti2004} & \citet{Gorski2011} \\
    \hline
	\end{tabular}
	\end{center}
    \tablecomments{Measured distance moduli for Sculptor using RRL variables and the TRGB. Since distance moduli depend heavily on the calibration used, all calibration references have been listed. MV15 values use their averaged, semi-empirical measurements, which does not report a systematic error.}
    \end{rotatetable}
\end{table*}

\section{\label{sec:Conclusion} \textbf{Summary}}

We have determined new, high-precision, independent Population II distances to the Sculptor dwarf spheroidal galaxy, a member of the Local Group.

Adopting a TRGB zero-point calibration from \citet{Freedman2020}, we find a TRGB distance modulus of $\mu^\mathrm{TRGB} = 19.59 \pm 0.07_\mathrm{stat} \pm 0.05_\mathrm{sys}$ mag. Using periods from the literature \citep{Martinez-Vazquez2015}, we were able to accurately determine mean magnitudes of RRLs identified in our images. Using these values and a Galactic calibration from Gaia DR2 parallaxes \citep{Neeley2019}, we determine separate distances from each of the PL/PW and PLZ/PWZ relations. We adopt the PWZ distance modulus of $\mu_{W_{I,V-I}}^\mathrm{RRL} = 19.60 \pm 0.01_\mathrm{stat} \pm 0.05_\mathrm{sys}$ mag. Finally, based on a non-parametric fiducial HB, we determine a HB distance modulus of $\mu^\mathrm{HB} = 19.54 \pm 0.03_\mathrm{stat} \pm 0.09_\mathrm{sys}$ mag. We find that our results are in agreement with previous literature measurements using similar methods. In particular, we find that the three distance estimates we have determined are consistent with each other to within one standard deviation.

\acknowledgements

Based, in part, on observations taken with the IMACS camera on Baade/Magellan 6.5m telescope at the Las Campanas Observatories, which is owned and operated by the Carnegie Institution for Science.

Support for program \#13691 was provided by NASA through a grant from the Space Telescope Science Institute, which is operated by the Association of Universities for Research in Astronomy, Inc., under NASA contract NASA 5-26555.

This research has made use of the NASA/IPAC Extragalactic Database (NED), which is operated by the Jet Propulsion Laboratory, California Institute of Technology, under contract with the National Aeronautics and Space Administration.
Some of the data presented in this paper were obtained from the Mikulski Archive for Space Telescopes (MAST). STScI is operated by the Association of Universities for Research in Astronomy, Inc., under NASA contract NAS5-26555. 
We thank the {\it Observatories of the Carnegie Institution for Science} and the {\it University of Chicago} for their support of long-term research into the calibration and determination of the local expansion rate of the Universe in the more-distant Hubble flow.

This work has made use of data from the European Space Agency (ESA) mission {\it Gaia} (\url{https://www.cosmos.esa.int/gaia}), processed by the {\it Gaia} Data Processing and Analysis Consortium (DPAC,
\url{https://www.cosmos.esa.int/web/gaia/dpac/consortium}). Funding for the DPAC has been provided by national institutions, in particular the institutions participating in the {\it Gaia} Multilateral Agreement.

\facilities{\textit{HST} (ACS/WFC), Magellan-Baade (IMACS)}
\software{DAOPHOT \citep{Stetson1987}, ALLFRAME \citep{Stetson1994}, \texttt{Astrometry.net} \citep{Lang2010}, TinyTim \citep{Krist2011}, \texttt{astropy} \citep{Astropy2018}, \texttt{matplotlib} \citep{Hunter4160265}}

\clearpage
\bibliography{Sculptor}

\begin{thebibliography}{}
\expandafter\ifx\csname natexlab\endcsname\relax\def\natexlab#1{#1}\fi
\providecommand{\url}[1]{\href{#1}{#1}}
\providecommand{\dodoi}[1]{doi:~\href{http://doi.org/#1}{\nolinkurl{#1}}}
\providecommand{\doeprint}[1]{\href{http://ascl.net/#1}{\nolinkurl{http://ascl.net/#1}}}
\providecommand{\doarXiv}[1]{\href{https://arxiv.org/abs/#1}{\nolinkurl{https://arxiv.org/abs/#1}}}

\bibitem[{{Astropy Collaboration} {et~al.}(2018){Astropy Collaboration},
  {Price-Whelan}, {Sip{\H{o}}cz}, {G{\"u}nther}, {Lim}, {Crawford}, {Conseil},
  {Shupe}, {Craig}, {Dencheva}, {Ginsburg}, {Vand erPlas}, {Bradley},
  {P{\'e}rez-Su{\'a}rez}, {de Val-Borro}, {Aldcroft}, {Cruz}, {Robitaille},
  {Tollerud}, {Ardelean}, {Babej}, {Bach}, {Bachetti}, {Bakanov}, {Bamford},
  {Barentsen}, {Barmby}, {Baumbach}, {Berry}, {Biscani}, {Boquien}, {Bostroem},
  {Bouma}, {Brammer}, {Bray}, {Breytenbach}, {Buddelmeijer}, {Burke},
  {Calderone}, {Cano Rodr{\'\i}guez}, {Cara}, {Cardoso}, {Cheedella}, {Copin},
  {Corrales}, {Crichton}, {D'Avella}, {Deil}, {Depagne}, {Dietrich}, {Donath},
  {Droettboom}, {Earl}, {Erben}, {Fabbro}, {Ferreira}, {Finethy}, {Fox},
  {Garrison}, {Gibbons}, {Goldstein}, {Gommers}, {Greco}, {Greenfield},
  {Groener}, {Grollier}, {Hagen}, {Hirst}, {Homeier}, {Horton}, {Hosseinzadeh},
  {Hu}, {Hunkeler}, {Ivezi{\'c}}, {Jain}, {Jenness}, {Kanarek}, {Kendrew},
  {Kern}, {Kerzendorf}, {Khvalko}, {King}, {Kirkby}, {Kulkarni}, {Kumar},
  {Lee}, {Lenz}, {Littlefair}, {Ma}, {Macleod}, {Mastropietro}, {McCully},
  {Montagnac}, {Morris}, {Mueller}, {Mumford}, {Muna}, {Murphy}, {Nelson},
  {Nguyen}, {Ninan}, {N{\"o}the}, {Ogaz}, {Oh}, {Parejko}, {Parley}, {Pascual},
  {Patil}, {Patil}, {Plunkett}, {Prochaska}, {Rastogi}, {Reddy Janga},
  {Sabater}, {Sakurikar}, {Seifert}, {Sherbert}, {Sherwood-Taylor}, {Shih},
  {Sick}, {Silbiger}, {Singanamalla}, {Singer}, {Sladen}, {Sooley},
  {Sornarajah}, {Streicher}, {Teuben}, {Thomas}, {Tremblay}, {Turner},
  {Terr{\'o}n}, {van Kerkwijk}, {de la Vega}, {Watkins}, {Weaver}, {Whitmore},
  {Woillez}, {Zabalza}, \& {Astropy Contributors}}]{Astropy2018}
{Astropy Collaboration}, {Price-Whelan}, A.~M., {Sip{\H{o}}cz}, B.~M., {et~al.}
  2018, \aj, 156, 123, \dodoi{10.3847/1538-3881/aabc4f}

\bibitem[{{Beaton} {et~al.}(2016){Beaton}, {Freedman}, {Madore}, {Bono},
  {Carlson}, {Clementini}, {Durbin}, {Garofalo}, {Hatt}, {Jang}, {Kollmeier},
  {Lee}, {Monson}, {Rich}, {Scowcroft}, {Seibert}, {Sturch}, \&
  {Yang}}]{Beaton2016}
{Beaton}, R.~L., {Freedman}, W.~L., {Madore}, B.~F., {et~al.} 2016, \apj, 832,
  210, \dodoi{10.3847/0004-637X/832/2/210}

\bibitem[{{Beaton} {et~al.}(2019){Beaton}, {Seibert}, {Hatt}, {Freedman},
  {Hoyt}, {Jang}, {Lee}, {Madore}, {Monson}, {Neeley}, {Rich}, \&
  {Scowcroft}}]{Beaton2019}
{Beaton}, R.~L., {Seibert}, M., {Hatt}, D., {et~al.} 2019, \apj, 885, 141,
  \dodoi{10.3847/1538-4357/ab4263}

\bibitem[{{Bellazzini} {et~al.}(2001){Bellazzini}, {Ferraro}, \&
  {Pancino}}]{Bellazzini2001}
{Bellazzini}, M., {Ferraro}, F.~R., \& {Pancino}, E. 2001, \apj, 556, 635,
  \dodoi{10.1086/321613}

\bibitem[{{Bernard} {et~al.}(2009){Bernard}, {Monelli}, {Gallart},
  {Drozdovsky}, {Stetson}, {Aparicio}, {Cassisi}, {Mayer}, {Cole}, {Hidalgo},
  {Skillman}, \& {Tolstoy}}]{Bernard2009}
{Bernard}, E.~J., {Monelli}, M., {Gallart}, C., {et~al.} 2009, \apj, 699, 1742,
  \dodoi{10.1088/0004-637X/699/2/1742}

\bibitem[{{Bettinelli} {et~al.}(2019){Bettinelli}, {Hidalgo}, {Cassisi},
  {Aparicio}, {Piotto}, {Valdes}, \& {Walker}}]{Bettinelli2019}
{Bettinelli}, M., {Hidalgo}, S.~L., {Cassisi}, S., {et~al.} 2019, \mnras, 487,
  5862, \dodoi{10.1093/mnras/stz1679}

\bibitem[{{Bono} {et~al.}(2003){Bono}, {Caputo}, {Castellani}, {Marconi},
  {Storm}, \& {Degl'Innocenti}}]{Bono2003}
{Bono}, G., {Caputo}, F., {Castellani}, V., {et~al.} 2003, MNRAS, 344, 1097,
  \dodoi{10.1046/j.1365-8711.2003.06878.x}

\bibitem[{{Braga} {et~al.}(2015){Braga}, {Dall'Ora}, {Bono}, {Stetson},
  {Ferraro}, {Iannicola}, {Marengo}, {Neeley}, {Persson}, {Buonanno},
  {Coppola}, {Freedman}, {Madore}, {Marconi}, {Matsunaga}, {Monson}, {Rich},
  {Scowcroft}, \& {Seibert}}]{Braga2015}
{Braga}, V.~F., {Dall'Ora}, M., {Bono}, G., {et~al.} 2015, \apj, 799, 165,
  \dodoi{10.1088/0004-637X/799/2/165}

\bibitem[{{Carretta} {et~al.}(2000){Carretta}, {Gratton}, {Clementini}, \&
  {Fusi Pecci}}]{Carretta2000}
{Carretta}, E., {Gratton}, R.~G., {Clementini}, G., \& {Fusi Pecci}, F. 2000,
  \apj, 533, 215, \dodoi{10.1086/308629}

\bibitem[{{Catelan} {et~al.}(2004){Catelan}, {Pritzl}, \&
  {Smith}}]{Catelan2004}
{Catelan}, M., {Pritzl}, B.~J., \& {Smith}, H.~A. 2004, APJS, 154, 633,
  \dodoi{10.1086/422916}

\bibitem[{{Cerny} {et~al.}(2020){Cerny}, {Freedman}, {Madore}, {Ashmead},
  {Hoyt}, {Oakes}, {Quang Hoang Tran}, \& {Moss}}]{Cerny_2021}
{Cerny}, W., {Freedman}, W.~L., {Madore}, B.~F., {et~al.} 2020, arXiv e-prints,
  arXiv:2012.09701.
\newblock \doarXiv{2012.09701}

\bibitem[{Clem {et~al.}(2008)Clem, VandenBerg, \& Stetson}]{Clem_2008}
Clem, J.~L., VandenBerg, D.~A., \& Stetson, P.~B. 2008, The Astronomical
  Journal, 135, 682, \dodoi{10.1088/0004-6256/135/2/682}

\bibitem[{{Clementini} {et~al.}(2005){Clementini}, {Ripepi}, {Bragaglia},
  {Martinez Fiorenzano}, {Held}, \& {Gratton}}]{Clementini2005}
{Clementini}, G., {Ripepi}, V., {Bragaglia}, A., {et~al.} 2005, MNRAS, 363,
  734, \dodoi{10.1111/j.1365-2966.2005.09478.x}

\bibitem[{{Da Costa} \& {Armandroff}(1990)}]{DaCosta1990}
{Da Costa}, G.~S., \& {Armandroff}, T.~E. 1990, \aj, 100, 162,
  \dodoi{10.1086/115500}

\bibitem[{{de Boer} {et~al.}(2011){de Boer}, {Tolstoy}, {Saha}, {Olsen},
  {Irwin}, {Battaglia}, {Hill}, {Shetrone}, {Fiorentino}, \&
  {Cole}}]{deBoer2011}
{de Boer}, T.~J.~L., {Tolstoy}, E., {Saha}, A., {et~al.} 2011, \aap, 528, A119,
  \dodoi{10.1051/0004-6361/201016398}

\bibitem[{{Dotter} {et~al.}(2010){Dotter}, {Sarajedini}, {Anderson},
  {Aparicio}, {Bedin}, {Chaboyer}, {Majewski}, {Mar{\'\i}n-Franch}, {Milone},
  {Paust}, {Piotto}, {Reid}, {Rosenberg}, \& {Siegel}}]{Dotter2010}
{Dotter}, A., {Sarajedini}, A., {Anderson}, J., {et~al.} 2010, \apj, 708, 698,
  \dodoi{10.1088/0004-637X/708/1/698}

\bibitem[{{Dressler} {et~al.}(2011){Dressler}, {Bigelow}, {Hare}, {Sutin},
  {Thompson}, {Burley}, {Epps}, {Oemler}, {Bagish}, {Birk}, {Clardy},
  {Gunnels}, {Kelson}, {Shectman}, \& {Osip}}]{Dressler2011}
{Dressler}, A., {Bigelow}, B., {Hare}, T., {et~al.} 2011, \pasp, 123, 288,
  \dodoi{10.1086/658908}

\bibitem[{{Fabricius} {et~al.}(2021){Fabricius}, {Luri}, {Arenou}, {Babusiaux},
  {Helmi}, {Muraveva}, {Reyl{\'e}}, {Spoto}, {Vallenari}, {Antoja}, {Balbinot},
  {Barache}, {Bauchet}, {Bragaglia}, {Busonero}, {Cantat-Gaudin}, {Carrasco},
  {Diakit{\'e}}, {Fabrizio}, {Figueras}, {Garcia-Gutierrez}, {Garofalo},
  {Jordi}, {Kervella}, {Khanna}, {Leclerc}, {Licata}, {Lambert}, {Marrese},
  {Masip}, {Ramos}, {Robichon}, {Robin}, {Romero-G{\'o}mez}, {Rubele}, \&
  {Weiler}}]{Fabricius_2021}
{Fabricius}, C., {Luri}, X., {Arenou}, F., {et~al.} 2021, Astronomy and
  Astrophysics, 649, A5, \dodoi{10.1051/0004-6361/202039834}

\bibitem[{{Federici} {et~al.}(2012){Federici}, {Cacciari}, {Bellazzini}, {Fusi
  Pecci}, {Galleti}, \& {Perina}}]{Federici_2012}
{Federici}, L., {Cacciari}, C., {Bellazzini}, M., {et~al.} 2012, \aap, 544,
  A155, \dodoi{10.1051/0004-6361/201219317}

\bibitem[{{Ferraro} {et~al.}(1999){Ferraro}, {Messineo}, {Fusi Pecci}, {de
  Palo}, {Straniero}, {Chieffi}, \& {Limongi}}]{1999AJ....118.1738F}
{Ferraro}, F.~R., {Messineo}, M., {Fusi Pecci}, F., {et~al.} 1999, \aj, 118,
  1738, \dodoi{10.1086/301029}

\bibitem[{{Fitzpatrick}(1999)}]{Fitzpatrick1999}
{Fitzpatrick}, E.~L. 1999, \pasp, 111, 63, \dodoi{10.1086/316293}

\bibitem[{{Freedman}(2014)}]{Freedman2014}
{Freedman}, W. 2014, {CHP-II: The Carnegie Hubble Program to Measure Ho to 3\%
  Using Population II}, HST Proposal

\bibitem[{{Freedman}(1988)}]{Freedman1988}
{Freedman}, W.~L. 1988, \aj, 96, 1248, \dodoi{10.1086/114878}

\bibitem[{{Freedman}(2021)}]{Freedman2021}
---. 2021, arXiv e-prints, arXiv:2106.15656.
\newblock \doarXiv{2106.15656}

\bibitem[{{Freedman} \& {Madore}(2010)}]{Freedman2010}
{Freedman}, W.~L., \& {Madore}, B.~F. 2010, \apj, 719, 335,
  \dodoi{10.1088/0004-637X/719/1/335}

\bibitem[{{Freedman} {et~al.}(2019){Freedman}, {Madore}, {Hatt}, {Hoyt},
  {Jang}, {Beaton}, {Burns}, {Lee}, {Monson}, {Neeley}, {Phillips}, {Rich}, \&
  {Seibert}}]{Freedman2019}
{Freedman}, W.~L., {Madore}, B.~F., {Hatt}, D., {et~al.} 2019, \apj, 882, 34,
  \dodoi{10.3847/1538-4357/ab2f73}

\bibitem[{{Freedman} {et~al.}(2020){Freedman}, {Madore}, {Hoyt}, {Jang},
  {Beaton}, {Lee}, {Monson}, {Neeley}, \& {Rich}}]{Freedman2020}
{Freedman}, W.~L., {Madore}, B.~F., {Hoyt}, T., {et~al.} 2020, \apj, 891, 57,
  \dodoi{10.3847/1538-4357/ab7339}

\bibitem[{{Gaia Collaboration} {et~al.}(2016){Gaia Collaboration}, {Prusti},
  {de Bruijne}, {Brown}, {Vallenari}, {Babusiaux}, {Bailer-Jones}, {Bastian},
  {Biermann}, {Evans}, {Eyer}, {Jansen}, {Jordi}, {Klioner}, {Lammers},
  {Lindegren}, {Luri}, {Mignard}, {Milligan}, {Panem}, {Poinsignon},
  {Pourbaix}, {Randich}, {Sarri}, {Sartoretti}, {Siddiqui}, {Soubiran},
  {Valette}, {van Leeuwen}, {Walton}, {Aerts}, {Arenou}, {Cropper}, {Drimmel},
  {H{\o}g}, {Katz}, {Lattanzi}, {O'Mullane}, {Grebel}, {Holland}, {Huc},
  {Passot}, {Bramante}, {Cacciari}, {Casta{\~n}eda}, {Chaoul}, {Cheek}, {De
  Angeli}, {Fabricius}, {Guerra}, {Hern{\'a}ndez}, {Jean-Antoine-Piccolo},
  {Masana}, {Messineo}, {Mowlavi}, {Nienartowicz}, {Ord{\'o}{\~n}ez-Blanco},
  {Panuzzo}, {Portell}, {Richards}, {Riello}, {Seabroke}, {Tanga},
  {Th{\'e}venin}, {Torra}, {Els}, {Gracia-Abril}, {Comoretto},
  {Garcia-Reinaldos}, {Lock}, {Mercier}, {Altmann}, {Andrae}, {Astraatmadja},
  {Bellas-Velidis}, {Benson}, {Berthier}, {Blomme}, {Busso}, {Carry},
  {Cellino}, {Clementini}, {Cowell}, {Creevey}, {Cuypers}, {Davidson}, {De
  Ridder}, {de Torres}, {Delchambre}, {Dell'Oro}, {Ducourant}, {Fr{\'e}mat},
  {Garc{\'\i}a-Torres}, {Gosset}, {Halbwachs}, {Hambly}, {Harrison}, {Hauser},
  {Hestroffer}, {Hodgkin}, {Huckle}, {Hutton}, {Jasniewicz}, {Jordan},
  {Kontizas}, {Korn}, {Lanzafame}, {Manteiga}, {Moitinho}, {Muinonen},
  {Osinde}, {Pancino}, {Pauwels}, {Petit}, {Recio-Blanco}, {Robin}, {Sarro},
  {Siopis}, {Smith}, {Smith}, {Sozzetti}, {Thuillot}, {van Reeven}, {Viala},
  {Abbas}, {Abreu Aramburu}, {Accart}, {Aguado}, {Allan}, {Allasia},
  {Altavilla}, {{\'A}lvarez}, {Alves}, {Anderson}, {Andrei}, {Anglada Varela},
  {Antiche}, {Antoja}, {Ant{\'o}n}, {Arcay}, {Atzei}, {Ayache}, {Bach},
  {Baker}, {Balaguer-N{\'u}{\~n}ez}, {Barache}, {Barata}, {Barbier}, {Barblan},
  {Baroni}, {Barrado y Navascu{\'e}s}, {Barros}, {Barstow}, {Becciani},
  {Bellazzini}, {Bellei}, {Bello Garc{\'\i}a}, {Belokurov}, {Bendjoya},
  {Berihuete}, {Bianchi}, {Bienaym{\'e}}, {Billebaud}, {Blagorodnova},
  {Blanco-Cuaresma}, {Boch}, {Bombrun}, {Borrachero}, {Bouquillon}, {Bourda},
  {Bouy}, {Bragaglia}, {Breddels}, {Brouillet}, {Br{\"u}semeister},
  {Bucciarelli}, {Budnik}, {Burgess}, {Burgon}, {Burlacu}, {Busonero}, {Buzzi},
  {Caffau}, {Cambras}, {Campbell}, {Cancelliere}, {Cantat-Gaudin}, {Carlucci},
  {Carrasco}, {Castellani}, {Charlot}, {Charnas}, {Charvet}, {Chassat},
  {Chiavassa}, {Clotet}, {Cocozza}, {Collins}, {Collins}, {Costigan}, {Crifo},
  {Cross}, {Crosta}, {Crowley}, {Dafonte}, {Damerdji}, {Dapergolas}, {David},
  {David}, {De Cat}, {de Felice}, {de Laverny}, {De Luise}, {De March}, {de
  Martino}, {de Souza}, {Debosscher}, {del Pozo}, {Delbo}, {Delgado},
  {Delgado}, {di Marco}, {Di Matteo}, {Diakite}, {Distefano}, {Dolding}, {Dos
  Anjos}, {Drazinos}, {Dur{\'a}n}, {Dzigan}, {Ecale}, {Edvardsson}, {Enke},
  {Erdmann}, {Escolar}, {Espina}, {Evans}, {Eynard Bontemps}, {Fabre},
  {Fabrizio}, {Faigler}, {Falc{\~a}o}, {Farr{\`a}s Casas}, {Faye}, {Federici},
  {Fedorets}, {Fern{\'a}ndez-Hern{\'a}ndez}, {Fernique}, {Fienga}, {Figueras},
  {Filippi}, {Findeisen}, {Fonti}, {Fouesneau}, {Fraile}, {Fraser}, {Fuchs},
  {Furnell}, {Gai}, {Galleti}, {Galluccio}, {Garabato}, {Garc{\'\i}a-Sedano},
  {Gar{\'e}}, {Garofalo}, {Garralda}, {Gavras}, {Gerssen}, {Geyer}, {Gilmore},
  {Girona}, {Giuffrida}, {Gomes}, {Gonz{\'a}lez-Marcos},
  {Gonz{\'a}lez-N{\'u}{\~n}ez}, {Gonz{\'a}lez-Vidal}, {Granvik}, {Guerrier},
  {Guillout}, {Guiraud}, {G{\'u}rpide}, {Guti{\'e}rrez-S{\'a}nchez}, {Guy},
  {Haigron}, {Hatzidimitriou}, {Haywood}, {Heiter}, {Helmi}, {Hobbs},
  {Hofmann}, {Holl}, {Holland}, {Hunt}, {Hypki}, {Icardi}, {Irwin}, {Jevardat
  de Fombelle}, {Jofr{\'e}}, {Jonker}, {Jorissen}, {Julbe}, {Karampelas},
  {Kochoska}, {Kohley}, {Kolenberg}, {Kontizas}, {Koposov}, {Kordopatis},
  {Koubsky}, {Kowalczyk}, {Krone-Martins}, {Kudryashova}, {Kull}, {Bachchan},
  {Lacoste-Seris}, {Lanza}, {Lavigne}, {Le Poncin-Lafitte}, {Lebreton},
  {Lebzelter}, {Leccia}, {Leclerc}, {Lecoeur-Taibi}, {Lemaitre}, {Lenhardt},
  {Leroux}, {Liao}, {Licata}, {Lindstr{\o}m}, {Lister}, {Livanou}, {Lobel},
  {L{\"o}ffler}, {L{\'o}pez}, {Lopez-Lozano}, {Lorenz}, {Loureiro},
  {MacDonald}, {Magalh{\~a}es Fernandes}, {Managau}, {Mann}, {Mantelet},
  {Marchal}, {Marchant}, {Marconi}, {Marie}, {Marinoni}, {Marrese},
  {Marschalk{\'o}}, {Marshall}, {Mart{\'\i}n-Fleitas}, {Martino}, {Mary},
  {Matijevi{\v{c}}}, {Mazeh}, {McMillan}, {Messina}, {Mestre}, {Michalik},
  {Millar}, {Miranda}, {Molina}, {Molinaro}, {Molinaro}, {Moln{\'a}r},
  {Moniez}, {Montegriffo}, {Monteiro}, {Mor}, {Mora}, {Morbidelli}, {Morel},
  {Morgenthaler}, {Morley}, {Morris}, {Mulone}, {Muraveva}, {Musella},
  {Narbonne}, {Nelemans}, {Nicastro}, {Noval}, {Ord{\'e}novic},
  {Ordieres-Mer{\'e}}, {Osborne}, {Pagani}, {Pagano}, {Pailler}, {Palacin},
  {Palaversa}, {Parsons}, {Paulsen}, {Pecoraro}, {Pedrosa}, {Pentik{\"a}inen},
  {Pereira}, {Pichon}, {Piersimoni}, {Pineau}, {Plachy}, {Plum}, {Poujoulet},
  {Pr{\v{s}}a}, {Pulone}, {Ragaini}, {Rago}, {Rambaux}, {Ramos-Lerate},
  {Ranalli}, {Rauw}, {Read}, {Regibo}, {Renk}, {Reyl{\'e}}, {Ribeiro},
  {Rimoldini}, {Ripepi}, {Riva}, {Rixon}, {Roelens}, {Romero-G{\'o}mez},
  {Rowell}, {Royer}, {Rudolph}, {Ruiz-Dern}, {Sadowski}, {Sagrist{\`a}
  Sell{\'e}s}, {Sahlmann}, {Salgado}, {Salguero}, {Sarasso}, {Savietto},
  {Schnorhk}, {Schultheis}, {Sciacca}, {Segol}, {Segovia}, {Segransan},
  {Serpell}, {Shih}, {Smareglia}, {Smart}, {Smith}, {Solano}, {Solitro},
  {Sordo}, {Soria Nieto}, {Souchay}, {Spagna}, {Spoto}, {Stampa}, {Steele},
  {Steidelm{\"u}ller}, {Stephenson}, {Stoev}, {Suess}, {S{\"u}veges}, {Surdej},
  {Szabados}, {Szegedi-Elek}, {Tapiador}, {Taris}, {Tauran}, {Taylor},
  {Teixeira}, {Terrett}, {Tingley}, {Trager}, {Turon}, {Ulla}, {Utrilla},
  {Valentini}, {van Elteren}, {Van Hemelryck}, {van Leeuwen}, {Varadi},
  {Vecchiato}, {Veljanoski}, {Via}, {Vicente}, {Vogt}, {Voss}, {Votruba},
  {Voutsinas}, {Walmsley}, {Weiler}, {Weingrill}, {Werner}, {Wevers},
  {Whitehead}, {Wyrzykowski}, {Yoldas}, {{\v{Z}}erjal}, {Zucker}, {Zurbach},
  {Zwitter}, {Alecu}, {Allen}, {Allende Prieto}, {Amorim},
  {Anglada-Escud{\'e}}, {Arsenijevic}, {Azaz}, {Balm}, {Beck}, {Bernstein},
  {Bigot}, {Bijaoui}, {Blasco}, {Bonfigli}, {Bono}, {Boudreault}, {Bressan},
  {Brown}, {Brunet}, {Bunclark}, {Buonanno}, {Butkevich}, {Carret}, {Carrion},
  {Chemin}, {Ch{\'e}reau}, {Corcione}, {Darmigny}, {de Boer}, {de Teodoro}, {de
  Zeeuw}, {Delle Luche}, {Domingues}, {Dubath}, {Fodor}, {Fr{\'e}zouls},
  {Fries}, {Fustes}, {Fyfe}, {Gallardo}, {Gallegos}, {Gardiol}, {Gebran},
  {Gomboc}, {G{\'o}mez}, {Grux}, {Gueguen}, {Heyrovsky}, {Hoar}, {Iannicola},
  {Isasi Parache}, {Janotto}, {Joliet}, {Jonckheere}, {Keil}, {Kim},
  {Klagyivik}, {Klar}, {Knude}, {Kochukhov}, {Kolka}, {Kos}, {Kutka}, {Lainey},
  {LeBouquin}, {Liu}, {Loreggia}, {Makarov}, {Marseille}, {Martayan},
  {Martinez-Rubi}, {Massart}, {Meynadier}, {Mignot}, {Munari}, {Nguyen},
  {Nordlander}, {Ocvirk}, {O'Flaherty}, {Olias Sanz}, {Ortiz}, {Osorio},
  {Oszkiewicz}, {Ouzounis}, {Palmer}, {Park}, {Pasquato}, {Peltzer}, {Peralta},
  {P{\'e}turaud}, {Pieniluoma}, {Pigozzi}, {Poels}, {Prat}, {Prod'homme},
  {Raison}, {Rebordao}, {Risquez}, {Rocca-Volmerange}, {Rosen}, {Ruiz-Fuertes},
  {Russo}, {Sembay}, {Serraller Vizcaino}, {Short}, {Siebert}, {Silva},
  {Sinachopoulos}, {Slezak}, {Soffel}, {Sosnowska}, {Strai{\v{z}}ys}, {ter
  Linden}, {Terrell}, {Theil}, {Tiede}, {Troisi}, {Tsalmantza}, {Tur},
  {Vaccari}, {Vachier}, {Valles}, {Van Hamme}, {Veltz}, {Virtanen}, {Wallut},
  {Wichmann}, {Wilkinson}, {Ziaeepour}, \& {Zschocke}}]{Gaia_2016}
{Gaia Collaboration}, {Prusti}, T., {de Bruijne}, J.~H.~J., {et~al.} 2016,
  Astronomy and Astrophysics, 595, A1, \dodoi{10.1051/0004-6361/201629272}

\bibitem[{{Gaia Collaboration} {et~al.}(2021){Gaia Collaboration}, {Brown},
  {Vallenari}, {Prusti}, {de Bruijne}, {Babusiaux}, {Biermann}, {Creevey},
  {Evans}, {Eyer}, {Hutton}, {Jansen}, {Jordi}, {Klioner}, {Lammers},
  {Lindegren}, {Luri}, {Mignard}, {Panem}, {Pourbaix}, {Randich}, {Sartoretti},
  {Soubiran}, {Walton}, {Arenou}, {Bailer-Jones}, {Bastian}, {Cropper},
  {Drimmel}, {Katz}, {Lattanzi}, {van Leeuwen}, {Bakker}, {Cacciari},
  {Casta{\~n}eda}, {De Angeli}, {Ducourant}, {Fabricius}, {Fouesneau},
  {Fr{\'e}mat}, {Guerra}, {Guerrier}, {Guiraud}, {Jean-Antoine Piccolo},
  {Masana}, {Messineo}, {Mowlavi}, {Nicolas}, {Nienartowicz}, {Pailler},
  {Panuzzo}, {Riclet}, {Roux}, {Seabroke}, {Sordo}, {Tanga}, {Th{\'e}venin},
  {Gracia-Abril}, {Portell}, {Teyssier}, {Altmann}, {Andrae}, {Bellas-Velidis},
  {Benson}, {Berthier}, {Blomme}, {Brugaletta}, {Burgess}, {Busso}, {Carry},
  {Cellino}, {Cheek}, {Clementini}, {Damerdji}, {Davidson}, {Delchambre},
  {Dell'Oro}, {Fern{\'a}ndez-Hern{\'a}ndez}, {Galluccio}, {Garc{\'\i}a-Lario},
  {Garcia-Reinaldos}, {Gonz{\'a}lez-N{\'u}{\~n}ez}, {Gosset}, {Haigron},
  {Halbwachs}, {Hambly}, {Harrison}, {Hatzidimitriou}, {Heiter},
  {Hern{\'a}ndez}, {Hestroffer}, {Hodgkin}, {Holl}, {Jan{\ss}en}, {Jevardat de
  Fombelle}, {Jordan}, {Krone-Martins}, {Lanzafame}, {L{\"o}ffler}, {Lorca},
  {Manteiga}, {Marchal}, {Marrese}, {Moitinho}, {Mora}, {Muinonen}, {Osborne},
  {Pancino}, {Pauwels}, {Petit}, {Recio-Blanco}, {Richards}, {Riello},
  {Rimoldini}, {Robin}, {Roegiers}, {Rybizki}, {Sarro}, {Siopis}, {Smith},
  {Sozzetti}, {Ulla}, {Utrilla}, {van Leeuwen}, {van Reeven}, {Abbas}, {Abreu
  Aramburu}, {Accart}, {Aerts}, {Aguado}, {Ajaj}, {Altavilla}, {{\'A}lvarez},
  {{\'A}lvarez Cid-Fuentes}, {Alves}, {Anderson}, {Anglada Varela}, {Antoja},
  {Audard}, {Baines}, {Baker}, {Balaguer-N{\'u}{\~n}ez}, {Balbinot}, {Balog},
  {Barache}, {Barbato}, {Barros}, {Barstow}, {Bartolom{\'e}}, {Bassilana},
  {Bauchet}, {Baudesson-Stella}, {Becciani}, {Bellazzini}, {Bernet}, {Bertone},
  {Bianchi}, {Blanco-Cuaresma}, {Boch}, {Bombrun}, {Bossini}, {Bouquillon},
  {Bragaglia}, {Bramante}, {Breedt}, {Bressan}, {Brouillet}, {Bucciarelli},
  {Burlacu}, {Busonero}, {Butkevich}, {Buzzi}, {Caffau}, {Cancelliere},
  {C{\'a}novas}, {Cantat-Gaudin}, {Carballo}, {Carlucci}, {Carnerero},
  {Carrasco}, {Casamiquela}, {Castellani}, {Castro-Ginard}, {Castro Sampol},
  {Chaoul}, {Charlot}, {Chemin}, {Chiavassa}, {Cioni}, {Comoretto}, {Cooper},
  {Cornez}, {Cowell}, {Crifo}, {Crosta}, {Crowley}, {Dafonte}, {Dapergolas},
  {David}, {David}, {de Laverny}, {De Luise}, {De March}, {De Ridder}, {de
  Souza}, {de Teodoro}, {de Torres}, {del Peloso}, {del Pozo}, {Delbo},
  {Delgado}, {Delgado}, {Delisle}, {Di Matteo}, {Diakite}, {Diener},
  {Distefano}, {Dolding}, {Eappachen}, {Edvardsson}, {Enke}, {Esquej}, {Fabre},
  {Fabrizio}, {Faigler}, {Fedorets}, {Fernique}, {Fienga}, {Figueras},
  {Fouron}, {Fragkoudi}, {Fraile}, {Franke}, {Gai}, {Garabato},
  {Garcia-Gutierrez}, {Garc{\'\i}a-Torres}, {Garofalo}, {Gavras}, {Gerlach},
  {Geyer}, {Giacobbe}, {Gilmore}, {Girona}, {Giuffrida}, {Gomel}, {Gomez},
  {Gonzalez-Santamaria}, {Gonz{\'a}lez-Vidal}, {Granvik},
  {Guti{\'e}rrez-S{\'a}nchez}, {Guy}, {Hauser}, {Haywood}, {Helmi}, {Hidalgo},
  {Hilger}, {H{\l}adczuk}, {Hobbs}, {Holland}, {Huckle}, {Jasniewicz},
  {Jonker}, {Juaristi Campillo}, {Julbe}, {Karbevska}, {Kervella}, {Khanna},
  {Kochoska}, {Kontizas}, {Kordopatis}, {Korn}, {Kostrzewa-Rutkowska},
  {Kruszy{\'n}ska}, {Lambert}, {Lanza}, {Lasne}, {Le Campion}, {Le Fustec},
  {Lebreton}, {Lebzelter}, {Leccia}, {Leclerc}, {Lecoeur-Taibi}, {Liao},
  {Licata}, {Lindstr{\o}m}, {Lister}, {Livanou}, {Lobel}, {Madrero Pardo},
  {Managau}, {Mann}, {Marchant}, {Marconi}, {Marcos Santos}, {Marinoni},
  {Marocco}, {Marshall}, {Martin Polo}, {Mart{\'\i}n-Fleitas}, {Masip},
  {Massari}, {Mastrobuono-Battisti}, {Mazeh}, {McMillan}, {Messina},
  {Michalik}, {Millar}, {Mints}, {Molina}, {Molinaro}, {Moln{\'a}r},
  {Montegriffo}, {Mor}, {Morbidelli}, {Morel}, {Morris}, {Mulone}, {Munoz},
  {Muraveva}, {Murphy}, {Musella}, {Noval}, {Ord{\'e}novic}, {Orr{\`u}},
  {Osinde}, {Pagani}, {Pagano}, {Palaversa}, {Palicio}, {Panahi}, {Pawlak},
  {Pe{\~n}alosa Esteller}, {Penttil{\"a}}, {Piersimoni}, {Pineau}, {Plachy},
  {Plum}, {Poggio}, {Poretti}, {Poujoulet}, {Pr{\v{s}}a}, {Pulone}, {Racero},
  {Ragaini}, {Rainer}, {Raiteri}, {Rambaux}, {Ramos}, {Ramos-Lerate}, {Re
  Fiorentin}, {Regibo}, {Reyl{\'e}}, {Ripepi}, {Riva}, {Rixon}, {Robichon},
  {Robin}, {Roelens}, {Rohrbasser}, {Romero-G{\'o}mez}, {Rowell}, {Royer},
  {Rybicki}, {Sadowski}, {Sagrist{\`a} Sell{\'e}s}, {Sahlmann}, {Salgado},
  {Salguero}, {Samaras}, {Sanchez Gimenez}, {Sanna}, {Santove{\~n}a},
  {Sarasso}, {Schultheis}, {Sciacca}, {Segol}, {Segovia}, {S{\'e}gransan},
  {Semeux}, {Shahaf}, {Siddiqui}, {Siebert}, {Siltala}, {Slezak}, {Smart},
  {Solano}, {Solitro}, {Souami}, {Souchay}, {Spagna}, {Spoto}, {Steele},
  {Steidelm{\"u}ller}, {Stephenson}, {S{\"u}veges}, {Szabados}, {Szegedi-Elek},
  {Taris}, {Tauran}, {Taylor}, {Teixeira}, {Thuillot}, {Tonello}, {Torra},
  {Torra}, {Turon}, {Unger}, {Vaillant}, {van Dillen}, {Vanel}, {Vecchiato},
  {Viala}, {Vicente}, {Voutsinas}, {Weiler}, {Wevers}, {Wyrzykowski}, {Yoldas},
  {Yvard}, {Zhao}, {Zorec}, {Zucker}, {Zurbach}, \& {Zwitter}}]{Gaia_2021}
{Gaia Collaboration}, {Brown}, A.~G.~A., {Vallenari}, A., {et~al.} 2021,
  Astronomy and Astrophysics, 649, A1, \dodoi{10.1051/0004-6361/202039657}

\bibitem[{{Garofalo} {et~al.}(2018){Garofalo}, {Scowcroft}, {Clementini},
  {Johnston}, {Cohen}, {Freedman}, {Madore}, {Majewski}, {Monson}, {Neeley},
  {Grillmair}, {Hendel}, {Kallivayalil}, {Marengo}, \& {van der
  Marel}}]{Garofalo2018}
{Garofalo}, A., {Scowcroft}, V., {Clementini}, G., {et~al.} 2018, \mnras, 481,
  578, \dodoi{10.1093/mnras/sty2222}

\bibitem[{{G{\'o}rski} {et~al.}(2011){G{\'o}rski}, {Pietrzy{\'n}ski}, \&
  {Gieren}}]{Gorski2011}
{G{\'o}rski}, M., {Pietrzy{\'n}ski}, G., \& {Gieren}, W. 2011, \aj, 141, 194,
  \dodoi{10.1088/0004-6256/141/6/194}

\bibitem[{G{\'{o}}rski {et~al.}(2018)G{\'{o}}rski, Pietrzy{\'{n}}ski, Gieren,
  Graczyk, Suchomska, Karczmarek, Cohen, Zgirski, Wielg{\'{o}}rski, Pilecki,
  Taormina, Ko{\l}aczkowski, \& Narloch}]{Gorski_2018}
G{\'{o}}rski, M., Pietrzy{\'{n}}ski, G., Gieren, W., {et~al.} 2018, The
  Astronomical Journal, 156, 278, \dodoi{10.3847/1538-3881/aaeacb}

\bibitem[{{Harris}(1996)}]{1996AJ....112.1487H}
{Harris}, W.~E. 1996, \aj, 112, 1487, \dodoi{10.1086/118116}

\bibitem[{{Hatt} {et~al.}(2017){Hatt}, {Beaton}, {Freedman}, {Madore}, {Jang},
  {Hoyt}, {Lee}, {Monson}, {Rich}, {Scowcroft}, \& {Seibert}}]{Hatt2017}
{Hatt}, D., {Beaton}, R.~L., {Freedman}, W.~L., {et~al.} 2017, \apj, 845, 146,
  \dodoi{10.3847/1538-4357/aa7f73}

\bibitem[{{Hill} {et~al.}(2019){Hill}, {Sk{\'u}lad{\'o}ttir}, {Tolstoy},
  {Venn}, {Shetrone}, {Jablonka}, {Primas}, {Battaglia}, {de Boer},
  {Fran{\c{c}}ois}, {Helmi}, {Kaufer}, {Letarte}, {Starkenburg}, \&
  {Spite}}]{Hill2019}
{Hill}, V., {Sk{\'u}lad{\'o}ttir}, {\'A}., {Tolstoy}, E., {et~al.} 2019, \aap,
  626, A15, \dodoi{10.1051/0004-6361/201833950}

\bibitem[{{Hoyt}(2021)}]{Hoyt2021}
{Hoyt}, T.~J. 2021, arXiv e-prints, arXiv:2106.13337.
\newblock \doarXiv{2106.13337}

\bibitem[{{Hunter}(2007)}]{Hunter4160265}
{Hunter}, J.~D. 2007, Computing in Science Engineering, 9, 90

\bibitem[{{Jang} \& {Lee}(2017)}]{Jang2017}
{Jang}, I.~S., \& {Lee}, M.~G. 2017, \apj, 835, 28,
  \dodoi{10.3847/1538-4357/835/1/28}

\bibitem[{{Kaluzny} {et~al.}(2014){Kaluzny}, {Thompson}, {Dotter}, {Rozyczka},
  {Pych}, {Rucinski}, \& {Burley}}]{2014AcA....64...11K}
{Kaluzny}, J., {Thompson}, I.~B., {Dotter}, A., {et~al.} 2014, \actaa, 64, 11.
\newblock \doarXiv{1403.6325}

\bibitem[{{Kirby} {et~al.}(2009){Kirby}, {Guhathakurta}, {Bolte}, {Sneden}, \&
  {Geha}}]{Kirby2009}
{Kirby}, E.~N., {Guhathakurta}, P., {Bolte}, M., {Sneden}, C., \& {Geha}, M.~C.
  2009, \apj, 705, 328, \dodoi{10.1088/0004-637X/705/1/328}

\bibitem[{{Krist} {et~al.}(2011){Krist}, {Hook}, \& {Stoehr}}]{Krist2011}
{Krist}, J.~E., {Hook}, R.~N., \& {Stoehr}, F. 2011, in Society of
  Photo-Optical Instrumentation Engineers (SPIE) Conference Series, Vol. 8127,
  Optical Modeling and Performance Predictions V, ed. M.~A. {Kahan}, 81270J,
  \dodoi{10.1117/12.892762}

\bibitem[{{Lang} {et~al.}(2010){Lang}, {Hogg}, {Mierle}, {Blanton}, \&
  {Roweis}}]{Lang2010}
{Lang}, D., {Hogg}, D.~W., {Mierle}, K., {Blanton}, M., \& {Roweis}, S. 2010,
  \aj, 139, 1782, \dodoi{10.1088/0004-6256/139/5/1782}

\bibitem[{{Lauberts}(1982)}]{Lauberts_1982}
{Lauberts}, A. 1982, {ESO/Uppsala survey of the ESO(B) atlas}

\bibitem[{{Layden}(1998)}]{Layden1998}
{Layden}, A.~C. 1998, \aj, 115, 193, \dodoi{10.1086/300195}

\bibitem[{{Lee} {et~al.}(1993){Lee}, {Freedman}, \& {Madore}}]{Lee1993}
{Lee}, M.~G., {Freedman}, W.~L., \& {Madore}, B.~F. 1993, \apj, 417, 553,
  \dodoi{10.1086/173334}

\bibitem[{{Lindegren} {et~al.}(2021){Lindegren}, {Klioner}, {Hern{\'a}ndez},
  {Bombrun}, {Ramos-Lerate}, {Steidelm{\"u}ller}, {Bastian}, {Biermann}, {de
  Torres}, {Gerlach}, {Geyer}, {Hilger}, {Hobbs}, {Lammers}, {McMillan},
  {Stephenson}, {Casta{\~n}eda}, {Davidson}, {Fabricius}, {Gracia-Abril},
  {Portell}, {Rowell}, {Teyssier}, {Torra}, {Bartolom{\'e}}, {Clotet},
  {Garralda}, {Gonz{\'a}lez-Vidal}, {Torra}, {Abbas}, {Altmann}, {Anglada
  Varela}, {Balaguer-N{\'u}{\~n}ez}, {Balog}, {Barache}, {Becciani}, {Bernet},
  {Bertone}, {Bianchi}, {Bouquillon}, {Brown}, {Bucciarelli}, {Busonero},
  {Butkevich}, {Buzzi}, {Cancelliere}, {Carlucci}, {Charlot}, {Cioni},
  {Crosta}, {Crowley}, {del Peloso}, {del Pozo}, {Drimmel}, {Esquej}, {Fienga},
  {Fraile}, {Gai}, {Garcia-Reinaldos}, {Guerra}, {Hambly}, {Hauser},
  {Jan{\ss}en}, {Jordan}, {Kostrzewa-Rutkowska}, {Lattanzi}, {Liao}, {Licata},
  {Lister}, {L{\"o}ffler}, {Marchant}, {Masip}, {Mignard}, {Mints}, {Molina},
  {Mora}, {Morbidelli}, {Murphy}, {Pagani}, {Panuzzo}, {Pe{\~n}alosa Esteller},
  {Poggio}, {Re Fiorentin}, {Riva}, {Sagrist{\`a} Sell{\'e}s}, {Sanchez
  Gimenez}, {Sarasso}, {Sciacca}, {Siddiqui}, {Smart}, {Souami}, {Spagna},
  {Steele}, {Taris}, {Utrilla}, {van Reeven}, \&
  {Vecchiato}}]{Lindegren_gaia_2021}
{Lindegren}, L., {Klioner}, S.~A., {Hern{\'a}ndez}, J., {et~al.} 2021,
  Astronomy and Astrophysics, 649, A2, \dodoi{10.1051/0004-6361/202039709}

\bibitem[{{Madore}(1982)}]{Madore1982}
{Madore}, B.~F. 1982, \apj, 253, 575, \dodoi{10.1086/159659}

\bibitem[{{Madore} {et~al.}(2009){Madore}, {Mager}, \& {Freedman}}]{Madore2009}
{Madore}, B.~F., {Mager}, V., \& {Freedman}, W.~L. 2009, \apj, 690, 389,
  \dodoi{10.1088/0004-637X/690/1/389}

\bibitem[{{Ma{\'\i}z Apell{\'a}niz} {et~al.}(2021){Ma{\'\i}z Apell{\'a}niz},
  {Pantaleoni Gonz{\'a}lez}, \& {Barb{\'a}}}]{Maiz_2021}
{Ma{\'\i}z Apell{\'a}niz}, J., {Pantaleoni Gonz{\'a}lez}, M., \& {Barb{\'a}},
  R.~H. 2021, arXiv e-prints, arXiv:2101.10206.
\newblock \doarXiv{2101.10206}

\bibitem[{{Majewski} {et~al.}(1999){Majewski}, {Siegel}, {Patterson}, \&
  {Rood}}]{Majewski_1999}
{Majewski}, S.~R., {Siegel}, M.~H., {Patterson}, R.~J., \& {Rood}, R.~T. 1999,
  \apjl, 520, L33, \dodoi{10.1086/312133}

\bibitem[{{Marconi} {et~al.}(2015){Marconi}, {Coppola}, {Bono}, {Braga},
  {Pietrinferni}, {Buonanno}, {Castellani}, {Musella}, {Ripepi}, \&
  {Stellingwerf}}]{Marconi2015}
{Marconi}, M., {Coppola}, G., {Bono}, G., {et~al.} 2015, \apj, 808, 50,
  \dodoi{10.1088/0004-637X/808/1/50}

\bibitem[{{Mart{\'{\i}}nez-V{\'a}zquez}
  {et~al.}(2015){Mart{\'{\i}}nez-V{\'a}zquez}, {Monelli}, {Bono}, {Stetson},
  {Ferraro}, {Bernard}, {Gallart}, {Fiorentino}, {Iannicola}, \&
  {Udalski}}]{Martinez-Vazquez2015}
{Mart{\'{\i}}nez-V{\'a}zquez}, C.~E., {Monelli}, M., {Bono}, G., {et~al.} 2015,
  MNRAS, 454, 1509, \dodoi{10.1093/mnras/stv2014}

\bibitem[{{Monson} {et~al.}(2017){Monson}, {Beaton}, {Scowcroft}, {Freedman},
  {Madore}, {Rich}, {Seibert}, {Kollmeier}, \& {Clementini}}]{Monson2017}
{Monson}, A.~J., {Beaton}, R.~L., {Scowcroft}, V., {et~al.} 2017, AJ, 153, 96,
  \dodoi{10.3847/1538-3881/153/3/96}

\bibitem[{{Muraveva} {et~al.}(2018){Muraveva}, {Garofalo}, {Scowcroft},
  {Clementini}, {Freedman}, {Madore}, \& {Monson}}]{Muraveva2018}
{Muraveva}, T., {Garofalo}, A., {Scowcroft}, V., {et~al.} 2018, \mnras, 480,
  4138, \dodoi{10.1093/mnras/sty1959}

\bibitem[{{Neeley} {et~al.}(2017){Neeley}, {Marengo}, {Bono}, {Braga},
  {Dall'Ora}, {Magurno}, {Marconi}, {Trueba}, {Tognelli}, {Prada Moroni},
  {Beaton}, {Freedman}, {Madore}, {Monson}, {Scowcroft}, {Seibert}, \&
  {Stetson}}]{Neeley2017}
{Neeley}, J.~R., {Marengo}, M., {Bono}, G., {et~al.} 2017, \apj, 841, 84,
  \dodoi{10.3847/1538-4357/aa713d}

\bibitem[{{Neeley} {et~al.}(2019){Neeley}, {Marengo}, {Freedman}, {Madore},
  {Beaton}, {Hatt}, {Hoyt}, {Monson}, {Rich}, {Sarajedini}, {Seibert}, \&
  {Scowcroft}}]{Neeley2019}
{Neeley}, J.~R., {Marengo}, M., {Freedman}, W.~L., {et~al.} 2019, \mnras, 490,
  4254, \dodoi{10.1093/mnras/stz2814}

\bibitem[{{Nikolov} {et~al.}(1984){Nikolov}, {Buchantsova}, \&
  {Frolov}}]{Nikolov_1984}
{Nikolov}, N., {Buchantsova}, N., \& {Frolov}, M. 1984, {Mean light and colour
  (B-V and U-B) curves of 210 field RR Lyrae typestars.}

\bibitem[{{Oakes} {et~al.}(2022){Oakes}, {Hoyt}, {Freedman}, {Madore}, {Tran},
  {Cerny}, {Beaton}, \& {Seibert}}]{Oakes2022}
{Oakes}, E.~K., {Hoyt}, T.~J., {Freedman}, W.~L., {et~al.} 2022, \apj, 929,
  116, \dodoi{10.3847/1538-4357/ac5b07}

\bibitem[{{Persson} {et~al.}(2004){Persson}, {Madore}, {Krzemi{\'n}ski},
  {Freedman}, {Roth}, \& {Murphy}}]{Persson2004}
{Persson}, S.~E., {Madore}, B.~F., {Krzemi{\'n}ski}, W., {et~al.} 2004, \aj,
  128, 2239, \dodoi{10.1086/424934}

\bibitem[{{Piatek} {et~al.}(2006){Piatek}, {Pryor}, {Bristow}, {Olszewski},
  {Harris}, {Mateo}, {Minniti}, \& {Tinney}}]{Piatek_2006}
{Piatek}, S., {Pryor}, C., {Bristow}, P., {et~al.} 2006, \aj, 131, 1445,
  \dodoi{10.1086/499526}

\bibitem[{{Pietrzy{\'n}ski} {et~al.}(2008){Pietrzy{\'n}ski}, {Gieren},
  {Szewczyk}, {Walker}, {Rizzi}, {Bresolin}, {Kudritzki}, {Nalewajko}, {Storm},
  {Dall'Ora}, \& {Ivanov}}]{Pietrzynski2008}
{Pietrzy{\'n}ski}, G., {Gieren}, W., {Szewczyk}, O., {et~al.} 2008, AJ, 135,
  1993, \dodoi{10.1088/0004-6256/135/6/1993}

\bibitem[{{Putman} {et~al.}(2021){Putman}, {Zheng}, {Price-Whelan}, {Grcevich},
  {Johnson}, {Tollerud}, \& {Peek}}]{Putman_2021}
{Putman}, M.~E., {Zheng}, Y., {Price-Whelan}, A.~M., {et~al.} 2021, \apj, 913,
  53, \dodoi{10.3847/1538-4357/abe391}

\bibitem[{{Rizzi} {et~al.}(2007){Rizzi}, {Tully}, {Makarov}, {Makarova},
  {Dolphin}, {Sakai}, \& {Shaya}}]{Rizzi2007}
{Rizzi}, L., {Tully}, R.~B., {Makarov}, D., {et~al.} 2007, \apj, 661, 815,
  \dodoi{10.1086/516566}

\bibitem[{{Salaris} \& {Cassisi}(2005)}]{Salaris2005}
{Salaris}, M., \& {Cassisi}, S. 2005, {Evolution of Stars and Stellar
  Populations}

\bibitem[{{Salomon} {et~al.}(2020){Salomon}, {Ibata}, {Reyl{\'e}}, {Famaey},
  {Libeskind}, {McConnachie}, \& {Hoffman}}]{Salomon_2020}
{Salomon}, J.~B., {Ibata}, R., {Reyl{\'e}}, C., {et~al.} 2020, arXiv e-prints,
  arXiv:2012.09204.
\newblock \doarXiv{2012.09204}

\bibitem[{{Schlafly} \& {Finkbeiner}(2011)}]{Schlafly2011}
{Schlafly}, E.~F., \& {Finkbeiner}, D.~P. 2011, \apj, 737, 103,
  \dodoi{10.1088/0004-637X/737/2/103}

\bibitem[{{Schlegel} {et~al.}(1998){Schlegel}, {Finkbeiner}, \&
  {Davis}}]{Schlegel1998}
{Schlegel}, D.~J., {Finkbeiner}, D.~P., \& {Davis}, M. 1998, \apj, 500, 525,
  \dodoi{10.1086/305772}

\bibitem[{{Serenelli} {et~al.}(2017){Serenelli}, {Weiss}, {Cassisi}, {Salaris},
  \& {Pietrinferni}}]{Serenelli2017}
{Serenelli}, A., {Weiss}, A., {Cassisi}, S., {Salaris}, M., \& {Pietrinferni},
  A. 2017, \aap, 606, A33, \dodoi{10.1051/0004-6361/201731004}

\bibitem[{{Shapley}(1938)}]{Shapley1938}
{Shapley}, H. 1938, Harvard College Observatory Bulletin, 908, 1

\bibitem[{{Sollima} {et~al.}(2006){Sollima}, {Cacciari}, \&
  {Valenti}}]{Sollima2006}
{Sollima}, A., {Cacciari}, C., \& {Valenti}, E. 2006, MNRAS, 372, 1675,
  \dodoi{10.1111/j.1365-2966.2006.10962.x}

\bibitem[{{Stetson}(1987)}]{Stetson1987}
{Stetson}, P.~B. 1987, PASP, 99, 191, \dodoi{10.1086/131977}

\bibitem[{{Stetson}(1994)}]{Stetson1994}
---. 1994, PASP, 106, 250, \dodoi{10.1086/133378}

\bibitem[{{Stetson}(2000)}]{Stetson2000}
---. 2000, DAOPHOT-II User Manual, Private Communication

\bibitem[{{Stetson} {et~al.}(2019){Stetson}, {Pancino}, {Zocchi}, {Sanna}, \&
  {Monelli}}]{2019MNRAS.485.3042S}
{Stetson}, P.~B., {Pancino}, E., {Zocchi}, A., {Sanna}, N., \& {Monelli}, M.
  2019, \mnras, 485, 3042, \dodoi{10.1093/mnras/stz585}

\bibitem[{{Thompson} {et~al.}(2001){Thompson}, {Kaluzny}, {Pych}, {Burley},
  {Krzeminski}, {Paczy{\'n}ski}, {Persson}, \& {Preston}}]{2001AJ....121.3089T}
{Thompson}, I.~B., {Kaluzny}, J., {Pych}, W., {et~al.} 2001, \aj, 121, 3089,
  \dodoi{10.1086/321084}

\bibitem[{{Tolstoy} {et~al.}(2009){Tolstoy}, {Hill}, \& {Tosi}}]{Tolstoy2009}
{Tolstoy}, E., {Hill}, V., \& {Tosi}, M. 2009, \araa, 47, 371,
  \dodoi{10.1146/annurev-astro-082708-101650}

\bibitem[{{Tolstoy} {et~al.}(2004){Tolstoy}, {Irwin}, {Helmi}, {Battaglia},
  {Jablonka}, {Hill}, {Venn}, {Shetrone}, {Letarte}, {Cole}, {Primas},
  {Francois}, {Arimoto}, {Sadakane}, {Kaufer}, {Szeifert}, \&
  {Abel}}]{Tolstoy2004}
{Tolstoy}, E., {Irwin}, M.~J., {Helmi}, A., {et~al.} 2004, \apjl, 617, L119,
  \dodoi{10.1086/427388}

\bibitem[{{Torra} {et~al.}(2021){Torra}, {Casta{\~n}eda}, {Fabricius},
  {Lindegren}, {Clotet}, {Gonz{\'a}lez-Vidal}, {Bartolom{\'e}}, {Bastian},
  {Bernet}, {Biermann}, {Garralda}, {G{\'u}rpide}, {Lammers}, {Portell}, \&
  {Torra}}]{Torra_2021}
{Torra}, F., {Casta{\~n}eda}, J., {Fabricius}, C., {et~al.} 2021, Astronomy and
  Astrophysics, 649, A10, \dodoi{10.1051/0004-6361/202039637}

\bibitem[{{Valenti} {et~al.}(2004){Valenti}, {Ferraro}, \&
  {Origlia}}]{Valenti2004}
{Valenti}, E., {Ferraro}, F.~R., \& {Origlia}, L. 2004, \mnras, 354, 815,
  \dodoi{10.1111/j.1365-2966.2004.08249.x}

\bibitem[{{Vasiliev} \& {Baumgardt}(2021)}]{Vasiliev_2021}
{Vasiliev}, E., \& {Baumgardt}, H. 2021, arXiv e-prints, arXiv:2102.09568.
\newblock \doarXiv{2102.09568}

\bibitem[{{Zinn} \& {West}(1984)}]{Zinn1984}
{Zinn}, R., \& {West}, M.~J. 1984, \apjs, 55, 45, \dodoi{10.1086/190947}

\end{thebibliography}
\bibliographystyle{aasjournal}

\restartappendixnumbering

\appendix

\section{Gaia EDR3 Proper-motion Cleaning}
\label{sec:appen_a}

\begin{figure*}
    \centering
    \includegraphics[width=0.9\textwidth]{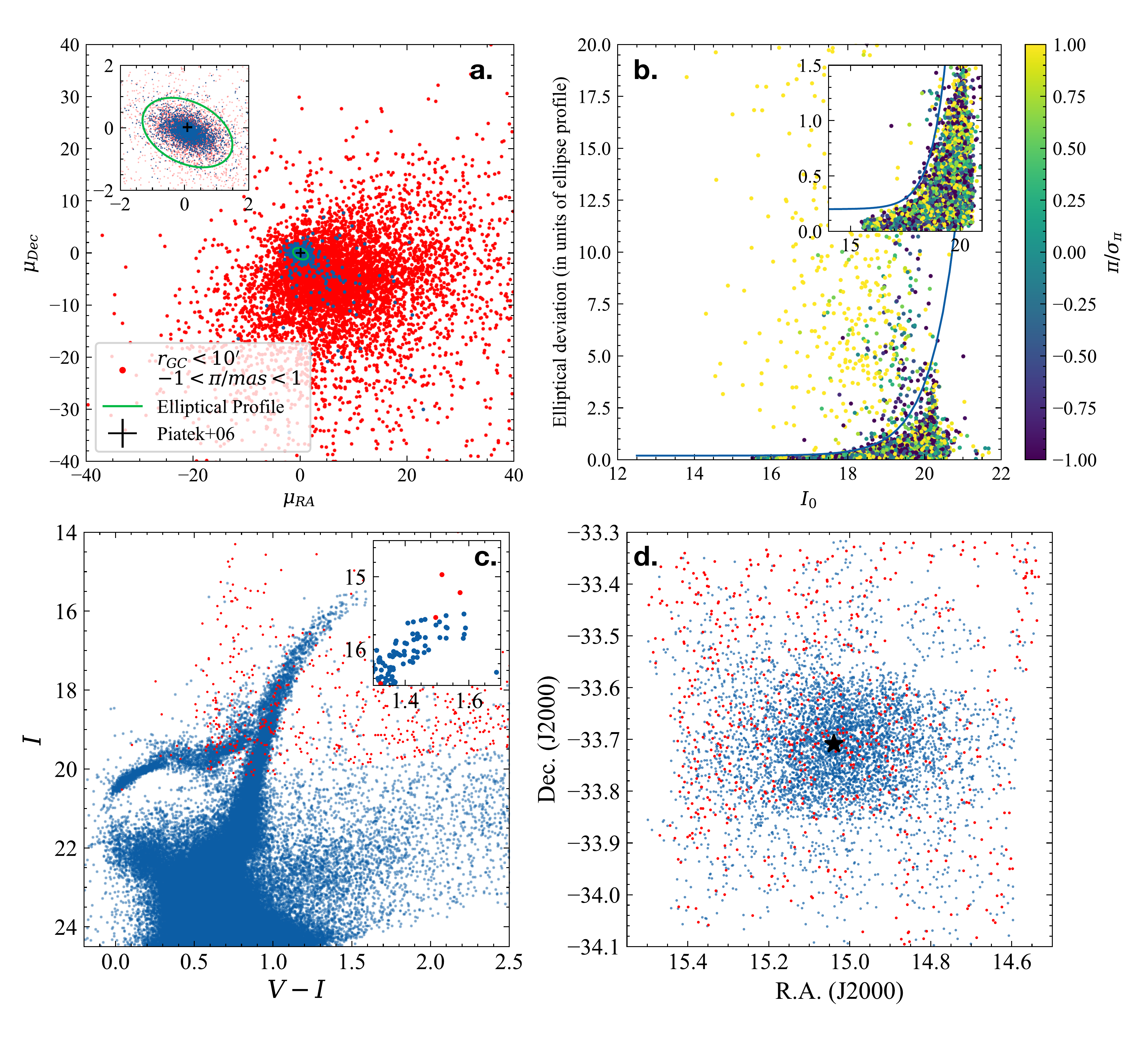}
    \caption{Gaia EDR3 foreground cleaning procedure. \textbf{a.} Proper motions of queried sources that satisfy the initial parallax and radial cut (blue) and those that do not (red). The Sculptor dSph proper motion determined by \citet{Piatek_2006} is plotted (black plus sign). \textbf{b.} Elliptical distance from the center of the adopted PM profile (green curve in panel a.) as a function of magnitude. Points are color coded according to their $\pi/\sigma_{\pi}$, to highlight the distribution of likely foreground sources (which will have large values of $\pi/\sigma_{\pi}$). The blue curve represents the envelope outside of which a source is classified as foreground. \textbf{c.} $I$ vs. $(V-I)$ CMD displaying the removed sources (red dots) and retained sources (blue dots) with a zoomed view of the TRGB shown in the inset. \textbf{d.} Spatial distribution of sources identified as foreground (red dots) and members of Sculptor (blue dots).}
    \label{fig:pmclean}
\end{figure*}

We selected all Gaia EDR3 sources within 1~deg of the Sculptor dSph centroid. The SQL query was,
\begin{verbatim}
SELECT * FROM gaiaedr3.gaia_source as edr3
WHERE 
1=CONTAINS(
	POINT('ICRS',edr3.ra,edr3.dec),
	CIRCLE(
		'ICRS',
		COORD1(EPOCH_PROP_POS(15.039166,-33.708888,
		0,.4760,-.3600,55.3000,2000,2016.0)),
		COORD2(EPOCH_PROP_POS(15.039166,-33.708888,
		0,.4760,-.3600,55.3000,2000,2016.0)),
		1.0)
)
\end{verbatim}
which returned 18060 sources.

We then made a cut for sources with parallaxes $-1 < \pi < 1 $~mas and that lie within $10'$ of Sculptor's optical center \citep[$01^h00^m09.4^s, -33^d42^m32^s$;][]{Lauberts_1982}. In \autoref{fig:pmclean}a, the distribution of sources that pass (blue) and don't (red) pass the selection criteria are plotted. These stars were then used to define an elliptical profile (simply fit ``by eye''). Also plotted is a determination of the Sculptor proper motion from HST imaging \citep[black cross,][]{Piatek_2006}. The parallax and radial cut reveals a tight cluster of sources in PM space centered near zero (emphasized in the inset). These sources were used to define an elliptical profile (green curve) that represents the 2D distribution of proper motions. The elliptical distance for each source was then computed, where a value of unity is equivalent to a coordinate $(\mu_{\alpha}, \mu_{\delta})$ located on the boundary of the adopted profile.

The PM elliptical distance was inspected as a function of magnitude (\autoref{fig:pmclean}b). It can be seen that fainter sources exhibit a larger scatter in PM space, within expectations. An exponential function was then adopted to describe the outer envelope of the elliptical deviation vs. $I$ distribution. 802 sources lie outside of the adopted envelope boundary and are conclusively identified as non-members of the Sculptor dwarf. The CMD and spatial distribution of sources (\autoref{fig:pmclean} c, d) confirm the accuracy of the proper-motion-cleaning procedure.

We briefly consider the possibility of a bias in the proper motions due to photometric color, for which \citet{Salomon_2020} found a sizable offset in the context of M31 (an order of magnitude more distant than Sculptor). Since the bright Sculptor CMD is dominated by red sources, while the faint Sculptor CMD by blue sources, it makes sense to check for this effect. We found for the blue sources (dominated by non-variable HB stars) $(\mu_{\alpha},\mu_{\delta}) = (0.128 \pm 0.013,-0.144 \pm 0.011)$~mas, and for the red sources $(\mu_{\alpha},\mu_{\delta}) = (0.100\pm 0.006,-0.153 \pm 0.004)$~mas, with only the formal standard errors (no systematic errors) quoted for both measurements. We see slight offsets between blue and red sources in both RA and DEC components of the proper motions, but the small difference has no effect on the foreground cleaning. While the subtle spatially covariant systematics of EDR3 astrometry ensure that our formal uncertainties are underestimated \citep[see, e.g.][]{Maiz_2021, Vasiliev_2021}, the consistency of Gaia EDR3 proper motions with an independent, dedicated HST program at a distance of 80~kpc is still quite remarkable.

\section{Comparison of Template Fits}
\label{sec:appen_b}

\begin{figure*}[!ht]
\centering
	\includegraphics[width=0.9\textwidth]{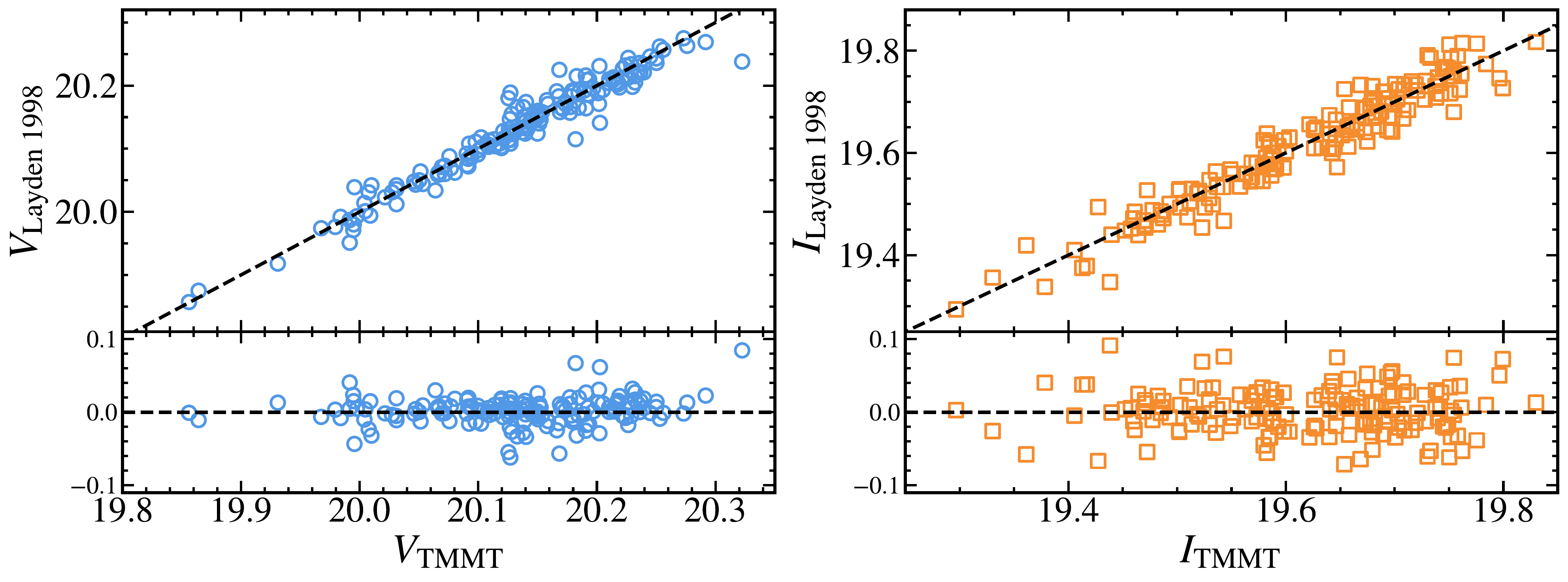}
    \caption{Comparison between TMMT template fits and MV15 templates \citep{Layden1998} for the $V$--band (blue circles, left) and $I$--band (orange squares, right) light curves from MV15. Residuals are shown in the bottom panel and the average and standard deviation of the offsets are $\Delta V = 0.002 \pm 0.019$ and $\Delta I = 0.002 \pm 0.031$ mag. The modern TMMT templates are shown to produce results in excellent agreement with those adopted by MV15, and with no functionality in the residuals.}
    \label{fig:template_compare}
\end{figure*}

To test the template-fitting procedure used in this work and confirm consistency with the MV15 template fits, the TMMT templates (described in \autoref{LC fits}) were fit to the MV15 light curves. 

To fit their light curves, MV15 used light curve templates partly based on the set used by \citet{Layden1998}, adopting the technique used in \citet{Bernard2009}. The \citet{Layden1998} templates for RRab stars were created by averaging and normalizing three to four well-sampled light curves taken from \citet{Nikolov_1984}.

The mean magnitudes from the TMMT fits to the MV15 light curves were then compared to MV15's own mean magnitudes. The results are shown in \autoref{fig:template_compare}. The differences between the TMMT-template mean magnitudes and the MV15 mean magnitudes are $\Delta V = 0.002 \pm 0.019$ and $\Delta I = 0.002 \pm 0.031$ mag, where the ``uncertainty'' quoted here is the standard deviation, not the standard error.

\section{\label{sec:HST_TRGB} Ground-to-\textit{HST} Transformation}
\label{sec:appen_c}

In this section we present a calibration between the space-based ACS/WFC VEGAMAG system to Stetson's Landolt system in the $I$--band. We used our ACS/WFC F814W images and repeated the procedure described in \autoref{sec:Photo} to match the calibrated Stetson $I$--band photometry with our \textit{HST} F814W photometry.

\begin{figure*}[!ht]
	\centering
	\includegraphics[width=1.0\textwidth]{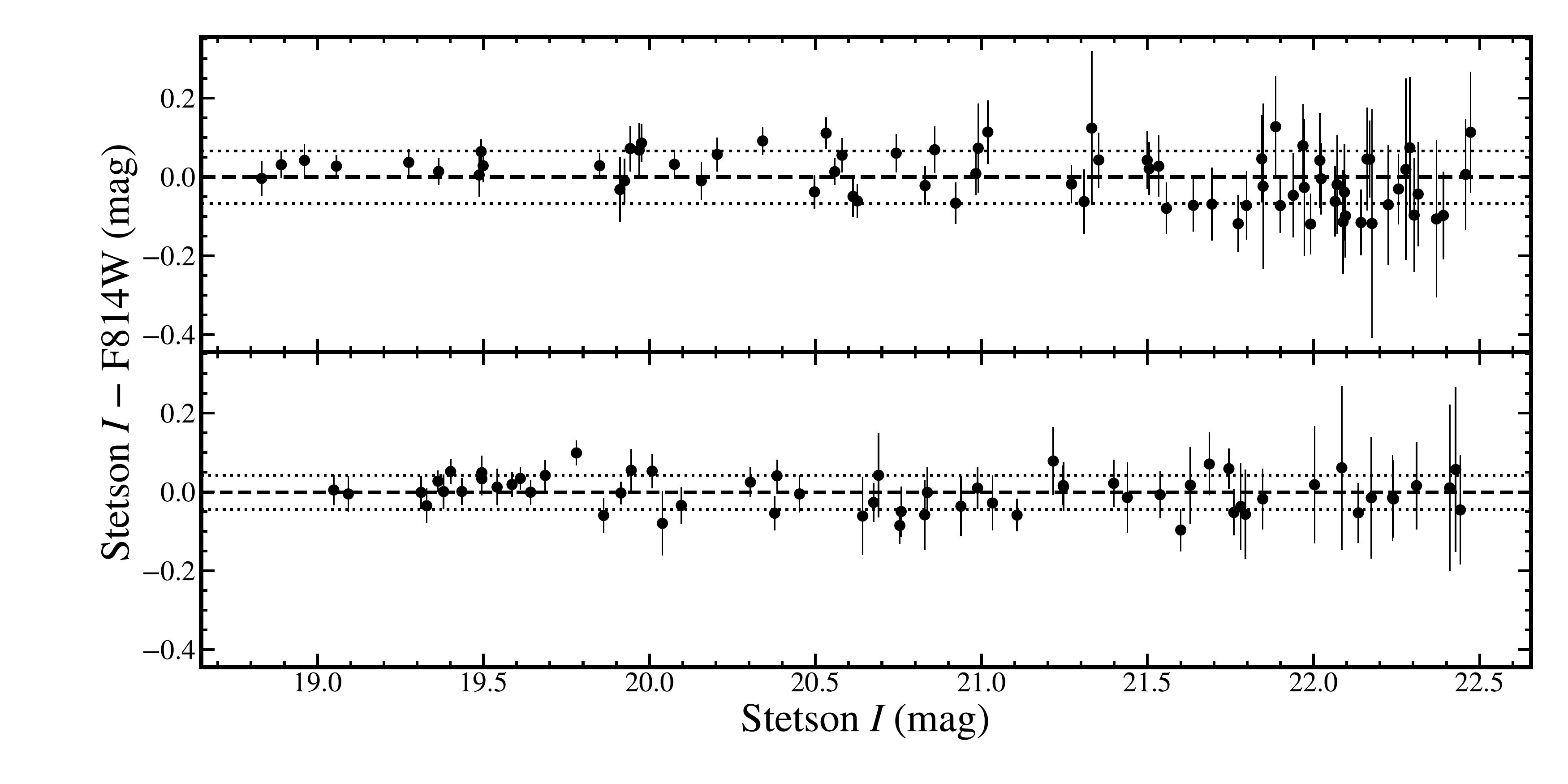}
	\caption{Comparison between Stetson $I$ and \textit{HST} F814W for the two IMACS pointings that overlap with the HST/ACS/F814W imaging. The top panel contains 73 comparison stars and the bottom 61 stars. The black dashed lines correspond to the mean offset and the black dotted lines are the 1$\sigma$ standard deviation. The top plot has a difference of $<0.001 \pm 0.066$ mag and the bottom plot has a difference of $-0.001 \pm 0.043$ mag between Stetson ground-based $I$--band and \textit{HST} F814W magnitudes.}%
	\label{fig:imacs_hst_comp}%
\end{figure*}

\autoref{fig:imacs_hst_comp} displays the calibration of Stetson $I$--band to the \textit{HST} F814W bandpass for 73 and 61 comparison stars from the two available different \textit{HST} pointings. The adopted mean Stetson $I$-F814W calibrations are plotted as dashed black lines and the 1$\sigma$ standard deviations as dotted black lines. Stars brighter than 18.8 mag in \textit{HST} imaging were removed due to concerns of non-linearity, while stars fainter than 22.5 mag in the IMACS imaging were removed to avoid potential false matches. For IMACS frames with overlapping \textit{HST} pointings, the average $I$--band and F814W offset is $<0.001 \pm 0.066$ mag and $-0.001 \pm 0.043$ mag while the average $I$--F814W offset is $<0.001 \pm 0.01$ mag. We find no evidence for a significant offset between the Stetson-Landolt $I$--band and ACS/F814W.

\end{document}